\documentclass[pre,twocolumn,nobalancelastpage,amsmath,amssymb,showpacs,
nofootinbib, superscriptaddress]{revtex4-2}

\usepackage{graphics}      
\usepackage{graphicx}      
\usepackage{longtable}     
\usepackage{url}           
\usepackage{bm}            
\usepackage{amsmath}
\usepackage{dcolumn}
\usepackage{mathtools}
\usepackage{hyperref}
\usepackage[dvipsnames]{xcolor}

\newcommand{\f}{\mathbf{f}}
\newcommand{\gdot}{\dot\gamma}

\newcommand{\bv}{\mathbf{v}}

\newcommand{\br}{\mathbf{r}}

\begin{document}
\title{Universality of stress-anisotropic and stress-isotropic jamming of frictionless spheres in three dimensions: Uniaxial vs isotropic compression }
\author{Anton Peshkov}
\affiliation{Department of Physics and Astronomy, University of Rochester, Rochester, NY 14627}
\author{S. Teitel}
\affiliation{Department of Physics and Astronomy, University of Rochester, Rochester, NY 14627}
\date{\today}

\begin{abstract}
We numerically study a three dimensional system of athermal, overdamped, frictionless spheres, using a simplified model for a non-Brownian suspension.  We compute the bulk viscosity under both uniaxial and isotropic compression as a means to address the question of whether stress-anisotropic and stress-isotropic jamming are in the same critical universality class.  Carrying out a critical scaling analysis of the system pressure $p$, shear stress $\sigma$, and macroscopic friction $\mu=\sigma/p$, as functions of particle packing fraction $\phi$ and compression rate $\dot\epsilon$, we find good agreement for all critical parameters comparing the isotropic and anisotropic cases.  In particular, we determine that the bulk viscosity diverges as $p/\dot\epsilon\sim (\phi_J-\phi)^{-\beta}$, with $\beta=3.36\pm 0.09$, as jamming is approached from below.  We further demonstrate that the average contact number per particle $Z$ can also be written in a scaling form as a function of $\phi$ and $\dot\epsilon$.  Once again, we find good agreement between the uniaxial and isotropic cases. 
We compare our results to prior simulations and theoretical predictions.
\end{abstract}
\maketitle


\section{Introduction}

Athermal ($T=0$) soft-core particles, with only contact interactions, have been widely used to model many soft-matter systems such as non-Brownian suspensions, emulsions, and foams.  Such systems undergo a jamming transition \cite{LiuNagel,OHern} as the particle packing fraction $\phi$ increases.  For packings $\phi$  below a critical $\phi_J$, the system behaves like a flowing fluid in response to small applied stresses; above $\phi_J$ the system behaves like a disordered but rigid solid, with a finite yield stress.  In this work we will consider a simple model for frictionless particles in suspension where, in the limit of small strain rates, the rheology of the fluid phase below jamming is Newtonian; the case of frictionless dry granular particles, where the fluid phase rheology is Bagnoldian, will be considered in a future work.  
For frictionless particles, as we consider here, the jamming transition is found to behave like a continuous phase transition \cite{OHern,OT1,OT2}, with transport coefficients  diverging continuously as $\phi\to\phi_J$ from below, and stress vanishing continuously to zero as $\phi\to\phi_J$ from above.

In the literature, jamming has been considered using several different protocols. (i) Random quenching  \cite{OHern,VOT}: in this protocol, initial configurations of randomly positioned soft-core particles are quenched at constant $\phi$ by rapid energy minimization.  In the large system limit, all configurations with $\phi<\phi_J$ will quench to zero energy, while those with $\phi>\phi_J$ will have finite energy. (ii) Isotropic quasistatic compression/decompression \cite{OHern,Chaudhuri}: in this protocol dilute comfigurations are isotropically and quasistatically compressed; the system box size is decreased in small discrete steps, with energy minimization of the configurations between subsequent steps.  In the large system limit, configurations with $\phi<\phi_J$ will have zero pressure, while above $\phi_J$ the pressure is finite.  In some versions of this protocol, the soft-core spheres are overcompressed above jamming, and then decompressed to determine $\phi_J$ as the point where the pressure drops to zero.  (iii) Shear-jamming \cite{Bi}: in the context of frictionless particles, shear-jamming occurs when an initially unjammed configuration of zero energy is  quasistatically simple-sheared with small discrete shear strain steps, with energy minimization between steps, until a mechanically stable configuration of finite energy is obtained \cite{Bertrand, Baity, Jin2}.  The initial configuration could either be one above $\phi_J$ \cite{Jin2,Vagberg.PRE.2011} (jammed states at $\phi_J$ are  considered to be random close packed, so unjammed configurations with a higher degree of order may continue to exist above $\phi_J$), or a configuration just below $\phi_J$, where shearing can induce a jammed state as a finite-size effect \cite{Bertrand,Baity}. (iv) Shear-driven jamming \cite{OT1,OT2,VOT,Hatano1,Hatano2,Hatano3,Otsuki,Heussinger1,Heussinger2}:  unlike quasistatic shear-jamming, shear-driven jamming refers to  systems driven at a finite constant simple-shear strain rate $\dot\gamma$, so as to create a flowing steady state. For $\phi<\phi_J$, the system particles flow no matter how small is $\dot\gamma$, and the shear viscosity diverges as $\phi\to\phi_J$ from below. For $\phi>\phi_J$, as $\dot\gamma\to 0$ the system flows with a finite yield stress that vanishes as $\phi\to\phi_J$ from above.  As the system flows at finite $\dot\gamma$, it passes though an ensemble of configurations that becomes independent of the starting configuration for long enough shearing.  Protocols (i), (ii), and (iii), in which the goal is to produce mechanically stable states, generally probe static structural properties of the jammed configurations.  In protocol (iv), however,  the applied strain rate introduces a control time scale which can be used to probe the dynamic behavior of the system as jamming is approached.  

Protocols (i) and (ii) we  refer to as stress-isotropic jamming.  The configurations produced have on average a zero shear stress, and the isotropic stress tensor is characterized solely by the system pressure.  Protocols (iii) and (iv) we  refer to as stress-anisotropic jamming.  The  shearing of the system, whether quasistatically or at a finite rate, results in a net shear stress in the system, and hence an anisotropic stress tensor.  Because symmetry is often a key factor in determining critical behavior  \cite{CL}, and as stress-isotropic and stress-anisotropic jamming produce states with different stress symmetry, one may wonder if these two cases  will be in the same critical universality class,  i.e. whether they are characterized by the same set of critical exponents that describe the divergence of viscosities and the vanishing of stress as $\phi_J$ is approached from below and from above, respectively.  The purpose of this work is to address this question.

Baity-Jesi et al. \cite{Baity} and Jin and Yoshino \cite{Jin2} have argued for a common universality by looking at the scaling of the static structural properties of pressure $p$ and average particle contact number $Z$ within the shear-jamming protocol (iii), and finding the same behaviors as found previously \cite{OHern,Wyart} for isotropic jamming in protocols (i) and (ii).  More recently, Ikeda et al. \cite{Ikeda} probed the dynamic behavior by considering  energy relaxation, using overdamped equations of motion to determine the global relaxation time $\tau$ upon approaching jamming from below, an approach originally taken by Olsson \cite{OlssonRelax}.  Comparing $\tau$ for isotropic random initial configurations vs that for anisotropic initial configurations obtained from simple sheared simulations as in protocol (iv), Ikeda et al. found a common  scaling relation of $\tau$ with the  contact number $Z$, thus arguing that isotropic and anisotropic jamming share a common universality also for dynamic behavior.  

However a subsequent work by Nishikawa et al. \cite{Nishikawa}, including several of the same authors as  Ref.~\cite{Ikeda}, challenged the conclusions of Ref.~\cite{Ikeda}; they claimed that when systems with a larger number of particles $N$ are considered, a surprising finite size dependence $\tau\sim N\ln N$ is found, and thus $\tau$ has no proper thermodynamic limit.  A more recent work by Olsson \cite{OlssonNew}, though, has critiqued the work of Nishikawa et al., challenging some of their conclusions but also pointing out other difficulties with using the relaxation time $\tau$ associated with global energy relaxation.

As an alternative to considering the problematic relaxation time $\tau$, in a recent letter \cite{PeshkovTeitel} we investigated the dynamic behavior of stress-isotropic jamming by simulating isotropic compression at finite compressive strain rates $\dot\epsilon$ \cite{Torq1}.
We found that the pressure $p$ and strain rate $\dot\epsilon$ obeyed a similar critical scaling relation as was found previously \cite{OT1,OT2} for the shear-driven jamming of protocol (iv), and that the bulk viscosity $p/\dot\epsilon$ diverged as $\phi\to\phi_J$ from below, a reflection of the diverging critical time scale at jamming.  Comparing the critical exponents  from this scaling analysis of isotropic compression with those found previously in the literature for simple shear-driven jamming, we found excellent agreement for two dimensional systems  \cite{OT2}.  However our results for three dimensional (3D) systems remained inconclusive, primarily due to a wide range of values reported for the 3D simple shearing exponents in the literature \cite{Lerner,DeGiuli,Berthier,Olsson3D}.  

In the present work we readdress the question of the critical universality of isotropic and anisotropic jamming in three dimensions.  Instead of comparing our isotropic compression exponents to those found in simple shearing, here we compare them to results from uniaxial compression, which similarly creates jammed configurations with an anisotropic stress tensor.  Comparing our results for uniaxial compression with new results for isotropic compression, we now find that all critical parameters agree between the two cases.  We thus conclude that stress-anisotropic and stress-isotropic jamming are indeed in the same critical universality class for dynamic behaviors.

The remainder of our paper is organized as follows.  In Sec.~\ref{MM} we present our model and numerical methods.  In Sec.~\ref{Results} we present our numerical results for the critical scaling of pressure $p$, shear stress $\sigma$, macroscopic friction $\mu=\sigma/p$, and average particle contact number $Z$, comparing uniaxial vs isotropic compression.  In Sec.~\ref{Discuss} we discuss our results and relate them to prior simulations and to theoretical predictions.

\section{Model and Methods}
\label{MM}

Our model, originally introduced by O'Hern et al. \cite{OHern}, is one that has been widely used in the literature.  It consists of athermal ($T=0$),  bidisperse, frictionless, soft-core spheres in three dimensions.  There are equal numbers of big and small spheres, with {\color{black}respective diameters $d_b$ and $d_s$ in the} ratio $d_b/d_s=1.4$.  {\color{black}In the following, the subscript ``$b$'' will denote the big particles, while the subscript ``$s$" will denote the small particles.}

\subsection{Equations of Motion}

 For particles with center of mass at positions $\br_i$, two particles will interact with a one-sided harmonic contact repulsion whenever they overlap.  The interaction potential is,
\begin{equation}
U(r_{ij})=\left\{
\begin{array}{cl}
\frac{1}{2}k_e\left(1-\dfrac{r_{ij}}{d_{ij}}\right)^2,&\quad r_{ij}<d_{ij}\\[10pt]
0,&\quad r_{ij}>d_{ij}
\end{array}
\right.
\end{equation}
where $k_e$ is a stiffness constant,  $r_{ij}=|\br_i-\br_j|$ is the distance between the particles and $d_{ij}=(d_i+d_j)/2$ is their average diameter.  The elastic force acting on particle $i$ due to its contact with $j$ is then,
\begin{equation}
\f^\mathrm{el}_{ij}=-\dfrac{dU(r_{ij})}{d\br_i}
\end{equation}
Particles also experience a dissipative force.  As a simplified model for particles in solution, we take the dissipative force on particle $i$ to be a viscous drag on the particle with respect to the local velocity of the suspending host medium \cite{OT1,OT2,Durian},
\begin{equation}
\f_i^\mathrm{dis} =-k_dV_i\left[\dfrac{d\br_i}{dt}-\bv_\mathrm{host}(\br_i)\right]
\end{equation}
where $k_d$ is a dissipative constant, $V_i$ is the volume of particle $i$, and $\bv_\mathrm{host}(\br)$ is the velocity of the host medium at position $\br$.

Particle motion is determined by Newton's equation,
\begin{equation}
m_i\dfrac{d^2\br_i}{dt^2}=\f_i^\mathrm{el}+\f_i^\mathrm{dis},\qquad \f_i^\mathrm{el}={\sum_j}^\prime \f_{ij}^\mathrm{el}
\end{equation}
where the sum is over all particles $j$ in contact with $i$, and $m_i$ is the mass of particle $i$.  We take particle masses to be proportional to their volume, $m_i\propto V_i$.  Because our particles are spherical and frictionless, we ignore particle rotations.

We note that our model lacks many forces that might be important in particular real physical suspensions, such as gravity,  hydrodynamic forces \cite{hydro}, lubrication forces \cite{lub1,lub2,lub3}, and inter-particle contact friction \cite{DST0,DST1,DST2,DST3,DST4,DST5,DST6}.  Our goal is not to provide a realistic model of any particular system, but rather to use a simplified, idealized, model that allows us to accurately simulate large  systems at low strain rates, so as to investigate the dynamic critical behavior associated with frictionless jamming.  Just as frictionless models have played an important theoretical role in the study of systems of dry granular particles \cite{Roux}, we use our simplified model of  a suspension in the spirit that it is useful to first understand the behavior of simple models before adding more realistic complexities.  Our model has been widely used in the literature, particularly for studying the response to simple shearing \cite{OT1,OT2,Lerner,DeGiuli,Berthier,Olsson3D,Durian,OT3,Heussinger1,During,Tewari,Andreotti,Vagberg.PRL.2014}.

\subsection{Compression}

Our particles are placed in a rectangular box with side lengths $L_x, L_y$ and $L_z$ centered at $\br=0$, so that particle coordinates lie within the range $r_{i\mu}\in[-L_\mu/2,L_\mu/2]$.
To compress our system we take the velocity of the host medium, $\bv_\mathrm{host}(\br)$, to be an affine compression  at a fixed strain rate $\dot\epsilon$, while  shrinking  the appropriate box lengths at the same rate.
For an isotropic compression we take,
\begin{equation}
\bv_\mathrm{host}(\br)=-\dot\epsilon\br,\quad\text{isotropic compression}
\end{equation}
and
\begin{equation}
\dfrac{dL_\mu}{dt}=-\dot\epsilon L_\mu,\quad\mu = x,y,z.
\end{equation}
For uniaxial compression along the $\mathbf{\hat z}$ direction we take,
\begin{equation}
\bv_\mathrm{host}(\br)=-\dot\epsilon (\br\cdot\mathbf{\hat z})\mathbf{\hat z},\quad\text{uniaxial compression}
\end{equation}
and
\begin{equation}
\dfrac{dL_z}{dt}=-\dot\epsilon L_z, \quad L_x=L_y=\text{constant.}
\end{equation}

After each step of the numerical integration of the equations of motion, where the box lengths and particle positions change to 
\begin{equation}
L_\mu(t-\Delta t)\to L_\mu(t)\quad\text{and} \quad\br_i(t-\Delta t)\to\br_i(t), 
\end{equation}
if a particle winds up outside the system box, for example if $r_{i\mu}(t)>L_\mu(t)/2$, its position is then mapped back inside the box using periodic boundary conditions, 
\begin{equation}
\br_i(t)\to \br_i^\prime(t)=\br_i(t)-L_\mu(t)\boldsymbol{\hat\mu}, 
\end{equation}
and its velocity $\bv_i=d\br_i/dt$ is mapped to, 
\begin{equation}
\bv_i\to \bv_i^\prime = \bv_i -\bv_\mathrm{host}(\br_i)+\bv_\mathrm{host}(\br_i^\prime), 
\end{equation}
so that the fluctuation of the particle velocity with respect to the host medium remains the same.

\subsection{Simulation Method}

The equations of motion presented above depend on three dimensionless parameters \cite{Vag}.  The first is the  particle packing fraction,
\begin{equation}
\phi = \dfrac{1}{V}\sum_i V_i,\quad V=L_xL_yL_z.
\end{equation}
As we integrate over time, the system is compressed, the box volume $V$ decreases, and the packing $\phi$ increases.

Defining  time scales characteristic of the elastic and dissipative forces \cite{Vag},
\begin{equation}
\tau_e =\sqrt{ \dfrac{m_sd_s^2}{k_e}}\quad\text{and}\quad\tau_d=\dfrac{m_s}{k_dV_s}
\end{equation}
the second dimensionless parameter is the quality factor,
\begin{equation}
Q=\dfrac{\tau_d}{\tau_e} = \dfrac{\sqrt{m_sk_e}}{k_dV_sd_s}
\end{equation}
which measures the relative strengths of the dissipative and elastic forces.  As $Q$ decreases, inertial effects decrease.   For $Q$ sufficiently small, behavior becomes independent of the particular value of $Q$ and one enters the overdamped limit corresponding to massless particles, $m_s\to 0$ \cite{Vag,VagbergOlssonTeitel}.  For our simulations we will use $Q=1$, which is sufficiently small to put us in this overdamped limit \cite{Vag}.   

In the overdamped limit, $m_s\to 0$, both $\tau_e$ and $\tau_d\to 0$, however we can  define a time scale\footnote{The prefactor of $3/2$ is a historical artifact from an earlier work, and has no particular physical significance.} that remains finite \cite{Vag},
\begin{equation}
\tau_0  = \frac{3}{2}\,\dfrac{\tau_e^2}{\tau_d}=\frac{3}{2}\dfrac{\tau_e}{Q}= \frac{3}{2}\,\dfrac{k_dV_sd_s^2}{k_e}.
\end{equation}
Our third dimensionless parameter is then,
\begin{equation}
\dot\epsilon\tau_0,\quad\text{the dimensionless strain rate.}
\end{equation}
Henceforth we will take our unit of length to be $d_s=1$, and our unit of time to be $\tau_0=1$.  Quoted values of $\dot\epsilon$ are therefore the same as $\dot\epsilon\tau_0$.  We consider strain rates spanning the range $\dot\epsilon=10^{-8.5}$ to $10^{-5}$.

We use LAMMPS \cite{lammps} to integrate the equations of motion, using a time step of $\Delta t/\tau_0=0.1$.  
Unless otherwise noted, we use $N=32768$ total particles.  From our prior work \cite{PeshkovTeitel}, this $N$ is large enough to avoid finite size effects for the range of $\phi$ and $\dot\epsilon$ we consider.
Our simulations start with an initial configuration at low packing $\phi_\mathrm{init}=0.2$, constructed as follows.  We place particles down one by one at random, but making sure that there are no particle overlaps; if an overlap occurs, we discard that particle and try again until all $N$ particles are placed in the box.  Unlike in our previous work  \cite{PeshkovTeitel}, we start the compression runs at each $\dot\epsilon$ from independently constructed configurations at the same $\phi_\mathrm{init}$, so as to be certain that there are no correlations among the configurations at different $\dot\epsilon$.  For each $\dot\epsilon$ we average our results over compressions starting from 20 independent initial configurations.

For our simulations of isotropic compression, we use a cubic box with $L_x=L_y=L_z$.  For uniaxial compression we start at $\phi_\mathrm{init}$ with a rectangular box with $L_x=L_y < L_z$, such that the box becomes roughly cubic by the time we have compressed to the jamming $\phi_J$.

\subsection{Stress}

As we compress the system, at each $\phi$ we measure the stress tensor for the configuration arising from the elastic forces \cite{OHern},
\begin{equation}
\mathbf{P}=\dfrac{1}{V}\sum_{i<j}\f_{ij}^\mathrm{el}\otimes(\br_i-\br_j).
\end{equation}
A dimensionless stress tensor can be defined as \cite{Vag},
\begin{equation}
\mathbf{p}=\dfrac{\tau_e^2d_s}{m_s}\,\mathbf{P}.
\end{equation}
The dimensionless pressure is then,
\begin{equation}
p=\dfrac{\langle p_{xx}\rangle +\langle p_{yy}\rangle + \langle p_{zz}\rangle}{3}
\end{equation}
where $\langle\cdots\rangle$ indicates an average over our 20 independent compression runs.  For isotropic compression we have $\langle p_{xx}\rangle=\langle p_{yy}\rangle=\langle p_{zz}\rangle = p$, and the shear stress vanishes.

For  uniaxial compression along the $\mathbf{\hat z}$ direction, we assume the average stress tensor has the form,
\begin{equation}
\langle\mathbf{p}\rangle=\left[
\begin{array}{ccc}
p-\sigma/2 & 0 & 0\\
0 & p-\sigma/2 & 0\\
0 & 0 & p+\sigma
\end{array}
\right]
\end{equation}
and so the dimensionless shear stress $\sigma$ is,
\begin{equation}
\sigma=\dfrac{ 2\langle p_{zz}\rangle -\langle p_{xx}\rangle - \langle p_{yy}\rangle}{3}.
\label{esig}
\end{equation}
For both isotropic and uniaxial compression, the  off-diagonal elements of $\langle\mathbf{p}\rangle$ vanish.

For our dynamics with viscous drag, the rheology of our system is Newtonian at small strain rates $\dot\epsilon$ below jamming.  We therefore define the  bulk viscosity as,
\begin{equation}
\zeta = p/\dot\epsilon
\end{equation}
while, for uniaxial compression, the shear viscosity is defined as,
\begin{equation}
\eta=\sigma/\dot\epsilon
\end{equation}

For uniaxial compression the macroscopic friction is defined as,
\begin{equation}
\mu = {\sigma}/{p}
\end{equation}
and $\mu$ is in general finite, even though in our model there is no microscopic contact friction between particles.
To compare $\mu$ for uniaxial compression vs simple shearing, we can generalize the above definition and take $\mu=(p_\mathrm{max}-p)/p$, where $p_\mathrm{max}$ is the maximal eigenvalue of the stress tensor $\langle\mathbf{p}\rangle$.

\subsection{Critical Scaling}

It has been demonstrated \cite{OT1,OT2} that the rheology of frictionless spheres undergoing simple shearing at a fixed shear strain rate, obeys a critical scaling equation similar to that of continuous equilibrium phase transitions.  In a recent letter \cite{PeshkovTeitel}, we demonstrated that a similar critical scaling applies when frictionless spheres are isotropically compressed at a fixed compression rate $\dot\epsilon$.  The scaling equation for pressure, which holds as one asymptotically approaches the jamming critical point $(\phi_J,\dot\epsilon\to 0)$, is to leading order,
\begin{equation}
p(\phi,\dot\epsilon) = \dot\epsilon^q\,f\left(\dfrac{\phi-\phi_J}{\dot\epsilon^{1/z\nu}}\right).
\label{escale}
\end{equation}
The consequences of this scaling equation are as follows.  Exactly at jamming, $\phi=\phi_J$, the rheology is nonlinear,
\begin{equation}
p\sim\dot\epsilon^q,\qquad\text{at $\phi=\phi_J$}.
\label{escaleq}
\end{equation}
For $\phi<\phi_J$ below jamming, as $\dot\epsilon\to 0$, the rheology is linear $p\sim\dot\epsilon$  and $\zeta=p/\dot\epsilon$ approaches a finite limit.  This implies that $f(x\to -\infty)\sim |x|^{-(1-q)z\nu}$, and so,
\begin{equation}
\lim_{\dot\epsilon\to 0}\,\zeta \sim (\phi_J-\phi)^{-\beta},\quad\beta=(1-q)z\nu.
\label{escalebeta}
\end{equation}
The bulk viscosity $\zeta$ diverges algebraically with the exponent $\beta$ as one approaches jamming from below.

For $\phi>\phi_J$ above jamming, as $\dot\epsilon\to 0$, the pressure approaches a finite limit in the jammed solid.  This implies that $f(x\to+\infty)\sim x^{qz\nu}$, and so,
\begin{equation}
\lim_{\dot\epsilon\to 0}\,p\sim(\phi-\phi_J)^y,\quad y=qz\nu.
\label{escaley}
\end{equation}
The pressure increases algebraically from zero with exponent $y$, as the soft spheres are compressed above jamming.

Note, the exponent $\beta$ is expected to be independent of the specific form of the elastic contact interaction since it describes behavior in the $\dot\epsilon\to 0$ hard-core limit \cite{OT3}.  The exponent $y$, however, will be sensitive to the power-law form of the contact interaction since all behavior above $\phi_J$ depends on particles having overlapping contacts, and so the soft-core nature of the particle contact potential necessarily determines the scaling of the pressure  \cite{OHern}.  A review of scaling in the context of the shear-driven jamming transition may be found in Ref.~\cite{VagbergOlssonTeitel}.

We expect similar scaling equations will hold for $p$, and also for $\sigma$, when the system is uniaxially compressed.  Since $p$ and $\sigma$ are parts of the same stress tensor, we expect their scaling exponents will be equal, and this has been demonstrated to be the case for the rheology in simple shearing \cite{OT2}.  This implies that the macroscopic friction $\mu=\sigma/p$ should approach a finite limit as $\dot\epsilon\to 0$.   

\section{Results}
\label{Results}

\subsection{Scaling of Pressure}

\begin{figure}
\centering
\includegraphics[width=3.4in]{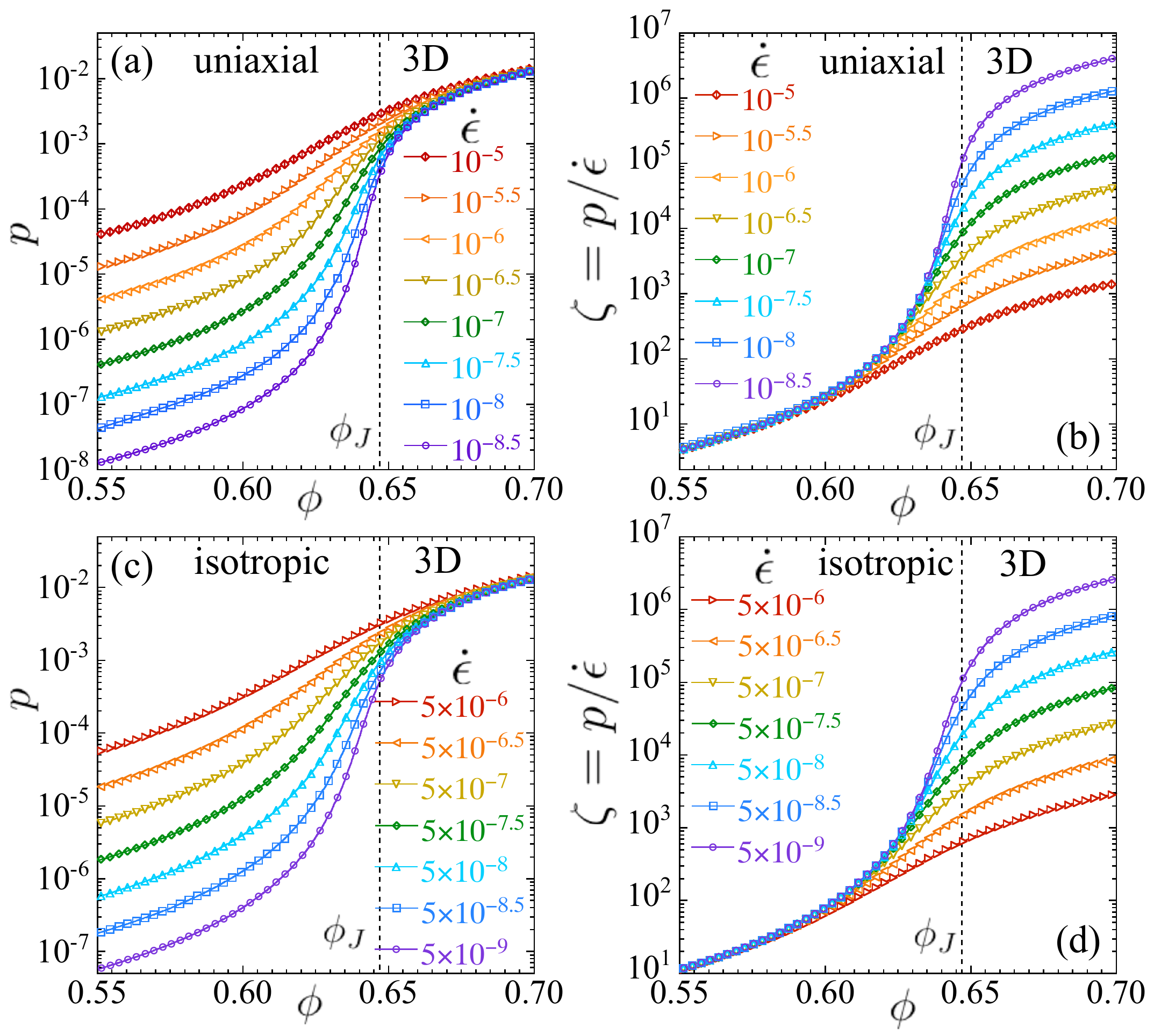}
\caption{(a) Pressure $p$ and (b) bulk viscosity $\zeta=p/\dot\epsilon$ vs packing $\phi$ for the uniaxial compression of bidisperse frictionless spheres in three dimension at different strain rates $\dot\epsilon$; (c) $p$ and (d) $\zeta$ for isotropic compression.  The vertical dashed lines locate the jamming $\phi_J$.  The system has $N=32768$ particles and results are averaged over 20 independent samples.  Error bars are roughly equal to or smaller than the size of the data symbols.
}
\label{f1}
\end{figure}

In Fig.~\ref{f1} we show our numerical results for the system pressure.  Figs.~\ref{f1}(a) and \ref{f1}(b) show pressure $p$ and bulk viscosity $\zeta=p/\dot\epsilon$ for uniaxial compression, while Figs.~\ref{f1}(c) and \ref{f1}(d) show $p$ and $\zeta$ for isotropic compression.  Note, we use a slightly different set of strain rates $\dot\epsilon$ for these two different cases.  

As predicted by  Eq.~(\ref{escaley}), we see that as $\dot\epsilon$ decreases, $p$ approaches a finite limit for $\phi>\phi_J$ but vanishes for $\phi<\phi_J$.  Similarly, as predicted by Eq.~(\ref{escalebeta}), we see that as $\dot\epsilon$ decreases, $\zeta$ approaches a finite limit for $\phi<\phi_J$ that appears to diverge as $\phi\to\phi_J$ from below.  Above $\phi_J$, where $p$ is finite,  $\zeta=p/\dot\epsilon$ diverges $\sim 1/\dot\epsilon$ as $\dot\epsilon\to 0$.

We now fit our data to the assumed scaling form.  Since the scaling function $f(x)$ of Eq.~(\ref{escale}) is not apriori known, we approximate it, for small values of its argument, by  the exponential of a fifth order polynomial,
\begin{equation}
f(x)=\mathrm{exp}\left({\sum_{n=0}^5 c_nx^n}\right).
\label{escalingf}
\end{equation}
We then fit our data to the form of Eq.~(\ref{escale}) regarding $\phi_J$, $q$, $1/z\nu$, and the polynomial coefficients $c_n$ as free fitting parameters.

The scaling equation (\ref{escale}) holds only asymptotically close to the critical point, i.e. as $\phi\to\phi_J$ and as $\dot\epsilon\to 0$.  One does not apriori know how close to the critical point one needs to be in order for the scaling to hold.  In order to test which of our data lies within the scaling region, we therefore fit  to Eq.~(\ref{escale}) using different windows of data, with $\phi\in[\phi_\mathrm{min},\phi_\mathrm{max}]$ and $\dot\epsilon\le\dot\epsilon_\mathrm{max}$.  If we find that our fitted parameters remain roughly constant within the estimated statistical error, as we shrink the data window, then we can have confidence that our fits are stable and self consistent.

\begin{figure}
\centering
\includegraphics[width=3.4in]{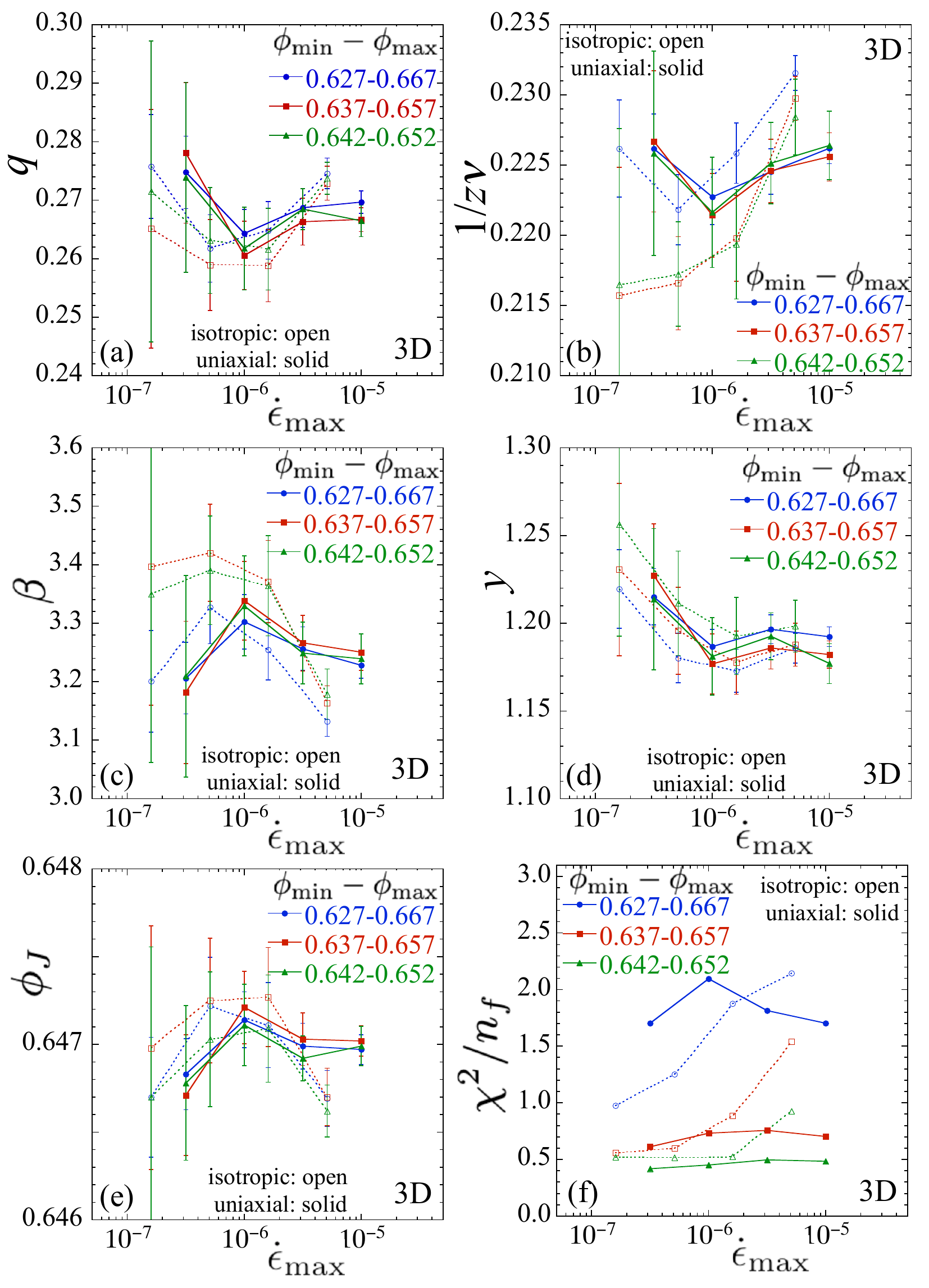}
\caption{Critical scaling parameters obtained by fitting the data of Fig.~\ref{f1} to the scaling form of Eq.~(\ref{escale}), vs the upper limit $\dot\epsilon_\mathrm{max}$ of the compression rate for data used in the fit.  Solid symbols and solid lines are for uniaxial compression while open symbols and dashed lines are for isotropic compression.  Results for three different ranges of packing $\phi\in[\phi_\mathrm{min},\phi_\mathrm{max}]$ are shown.  (a) exponent $q$; (b) exponent $1/z\nu$; (c) exponent $\beta=(1-q)z\nu$, (d) exponent $y=qz\nu$, (e) jamming $\phi_J$; (f) the $\chi^2$ per degree of freedom $n_f$ of the fits.  We use the jackknife method to compute the estimated errors and bias-corrected averages of the fit parameters.
}
\label{f2}
\end{figure}

In  Fig.~\ref{f2} we show the results from this fitting procedure, plotting the values of different fit parameters vs $\dot\epsilon_\mathrm{max}$, the maximum strain rate for data used in the fit.  We show results for three different windows of data $[\phi_\mathrm{min},\phi_\mathrm{max}]$ centered about the jamming $\phi_J$.
Solid data symbols connected by solid lines show our results for uniaxial compression, while open data symbols connected by dashed lines show our results for isotropic compression.  We use the jackknife method to estimate errors (one standard deviation statistical error) and bias-corrected averages of our fitting parameters. 
Figs.~\ref{f2}(a) and \ref{f2}(b) show the exponents $q$ and $1/z\nu$ that define the scaling equation (\ref{escale}).  Figs.~\ref{f2}(c) and \ref{f2}(d) show the related exponents $\beta=(1-q)z\nu$ and $y=qz\nu$ of Eqs.~(\ref{escalebeta}) and (\ref{escaley}).  Fig.~\ref{f2}(e) shows the jamming $\phi_J$ and \ref{f2}(f) shows the $\chi^2$ per degree of freedom $n_f$ of the fits.

We see that the fit parameters remain roughly equal, within the estimated errors, as $\dot\epsilon_\mathrm{max}$ decreases, and as the window $[\phi_\mathrm{min},\phi_\mathrm{max}]$ shrinks.  Moreover, we find that the critical parameters for uniaxial compression and isotropic compression are also equal within the estimated errors.  The $\chi^2/n_f$, shown in Fig.~\ref{f2}(f) shows that the quality of the fits improves as $[\phi_\mathrm{min},\phi_\mathrm{max}]$ shrinks, and for the smallest window of $\phi$ is roughly independent of $\dot\epsilon_\mathrm{max}$ for the smaller $\dot\epsilon$.  For the narrowest data window we have $\chi^2/n_f\approx 0.5$, suggesting a very good fit, however we caution that our compression protocol implies that (unlike in simple shearing) data points and their errors are strongly correlated as $\phi$ varies from one integration step to the next, and so the significance of the specific numerical value of $\chi^2/n_f$ is unclear.

We thus conclude that our fits are stable and self-consistent, and that in three dimensions stress-anisotropic jamming via uniaxial compression, and stress-isotropic jamming via isotropic compression, are in the same  critical universality class, characterized by the same rheological critical exponents.  This is our main conclusion.  Using our results from the smallest data window $\phi\in[0.642,0.653]$ and the second smallest $\dot\epsilon_\mathrm{max}$, we find,
\begin{equation} 
\begin{array}{rlrl}
q&=0.262\pm 0.009, &\quad 1/z\nu&=0.219\pm 0.004\\[10pt]
\beta&=3.36\pm0.09, &\quad y&=1.19\pm0.03.  
\end{array}
\label{eexps}
\end{equation}
We also find that the jamming
\begin{equation}\phi_J=0.6470\pm 0.0004
\label{ephiJ}
\end{equation} 
is the same for both uniaxial and isotropic compression.


\begin{figure}
\centering
\includegraphics[width=3.4in]{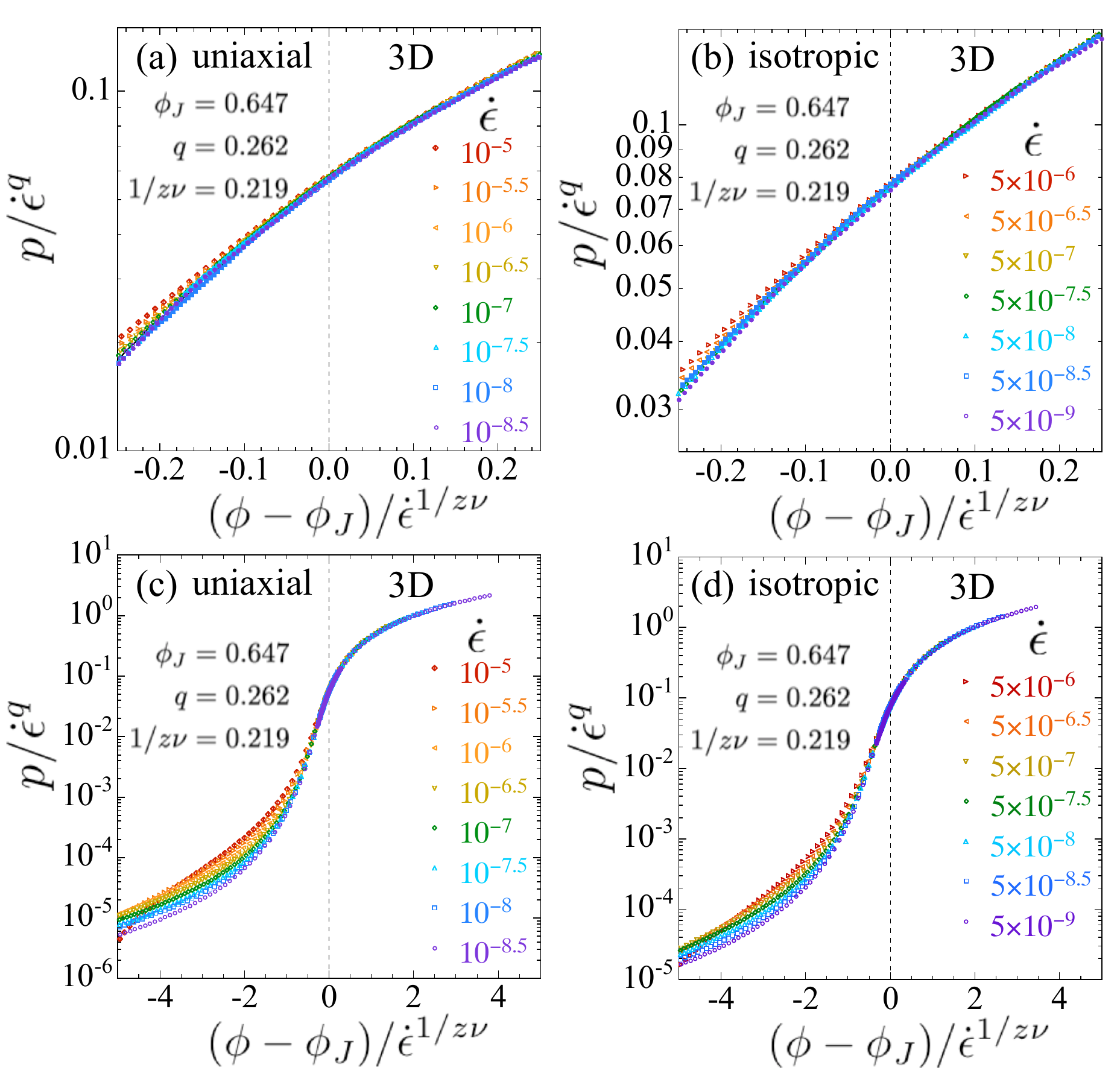}
\caption{Scaling collapses for the pressure data shown in Fig.~\ref{f1}, plotting $p/\dot\epsilon^q$ vs $x=(\phi-\phi_J)/\dot\epsilon^{1/z\nu}$ for (a,c) uniaxial compression and (b,d) isotropic compression.  The values of $q$, $1/z\nu$ and $\phi_J$, as indicated in the figure, are the same for both cases  and are obtained from the data window $\phi\in[0.642, 0.652]$ and the second smallest $\dot\epsilon_\mathrm{max}$.  The data points within this window are indicated by solid symbols, and span a narrow window with $|x|<0.23$; other data points are shown as open symbols.  {\color{black}(a) and (b) show an interval of $x$ that corresponds to the range of values used in making the fit.  (c) and (d) show a much wider range of $x$ that includes more of our data.}  We see a good collapse even for data that lie well outside the data window used in the fit.
}
\label{f3}
\end{figure}

As a further check of the critical scaling Eq.~(\ref{escale}), if we plot $p/\dot\epsilon^q$ vs $x\equiv (\phi-\phi_J)/\dot\epsilon^{1/z\nu}$ we expect our data for different compression rates $\dot\epsilon$ to collapse to a common scaling curve $f(x)$.  We show such a scaling plot in Fig.~\ref{f3}, using the values of the critical parameters given above in Eqs.~(\ref{eexps}) and (\ref{ephiJ}).  Fig.~\ref{f3}(a) is for uniaxial compression, while \ref{f3}(b) is for isotropic compression.  We see a generally good scaling collapse.  We note that the data used in the fit to the scaling equation {\color{black}(shown as solid symbols in Fig.~\ref{f3})} span a very narrow window with $|x|<0.23$.  {\color{black}In Figs.~\ref{f3}(c) and \ref{f3}(d) we show the same scaling plots, but now over a much wider range of $x$.  We see that the scaling collapse continues to hold over much of this wider range.} 
For small $x\lesssim -1$, below jamming, we see a departure from a common scaling curve for the larger values of $\dot\epsilon$; note also that for a fixed $x$, a larger value of $\dot\epsilon$ also implies a larger value of $|\phi-\phi_J|$.  We therefore believe this breakdown of a common scaling curve as $x$ decreases below $-1$ is due to the effect of corrections-to-scaling that become more significant as $\dot\epsilon$ increases, and $\phi$ decreases, and one goes further from the critical point $(\phi_J,\dot\epsilon\to 0)$.  We comment more on corrections-to-scaling in the next section.

In our previous letter on isotropic compression \cite{PeshkovTeitel}, we found for  three dimensions the critical parameters, $\phi_J=0.6464\pm 0.0005$, $\beta=3.07\pm 0.15$, and $y=1.22\pm 0.03$.  We note that those values of $\phi_J$ and $y$ are both within one standard deviation statistical error of the values found in the present work, while the value of $\beta$ is two standard deviations smaller.  It could be that the protocol we adopted in that earlier work, where simulations at each $\dot\epsilon$ were started from a configuration taken from simulations at the next higher value of $\dot\epsilon$, at successively larger values of $\phi_\mathrm{init}$, introduced   correlations between the data at different $\dot\epsilon$ that effected our analysis and led to small shifts in the fitted parameters.  It was because of this possibility that in the present work all simulations at different $\dot\epsilon$ were started from independent random configurations at the same $\phi_\mathrm{init}$.

\subsection{Scaling of Shear Stress}

\begin{figure}
\centering
\includegraphics[width=3.4in]{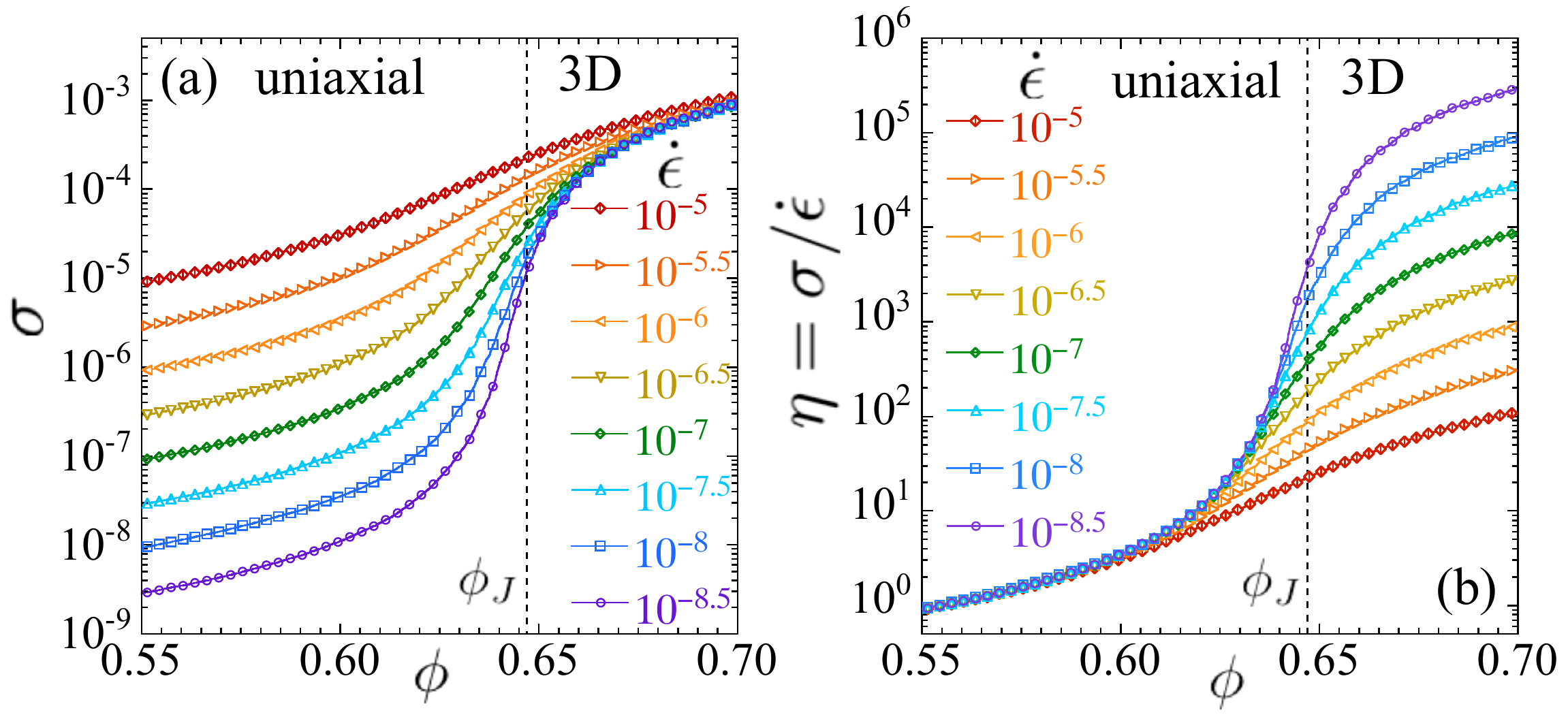}
\caption{(a) Shear stress $\sigma$ and (b) shear viscosity $\eta=\sigma/\dot\epsilon$ vs packing $\phi$ for the uniaxial compression of bidisperse frictionless spheres in three dimensions at different strain rates $\dot\epsilon$.  The system has $N=32768$ particles and results are averaged over 20 independent samples.  Error bars are roughly equal to or smaller than the size of the data symbols.  
}
\label{f4}
\end{figure}

In isotropic compression, the resulting average stress tensor is isotropic and completely characterized by the scalar pressure.  For uniaxial compression, however, we expect the system to develop a finite shear stress $\sigma$, as defined in Eq.~(\ref{esig}).  In Fig.~\ref{f4}(a) we plot the resulting $\sigma$ vs packing $\phi$, for different strain rates $\dot\epsilon$.  In Fig.~\ref{f4}(b) we plot the corresponding shear viscosity, $\eta=\sigma/\dot\epsilon$.  We see the same qualitative behavior as seen previously for the pressure $p$ in Fig.~\ref{f1}.  As $\dot\epsilon$ decreases, $\sigma$ approaches a finite limit for $\phi>\phi_J$ but vanishes for $\phi<\phi_J$, while $\eta$ approaches a finite limit for $\phi<\phi_J$ and diverges for $\phi>\phi_J$.

\begin{figure}
\centering
\includegraphics[height=1.8in]{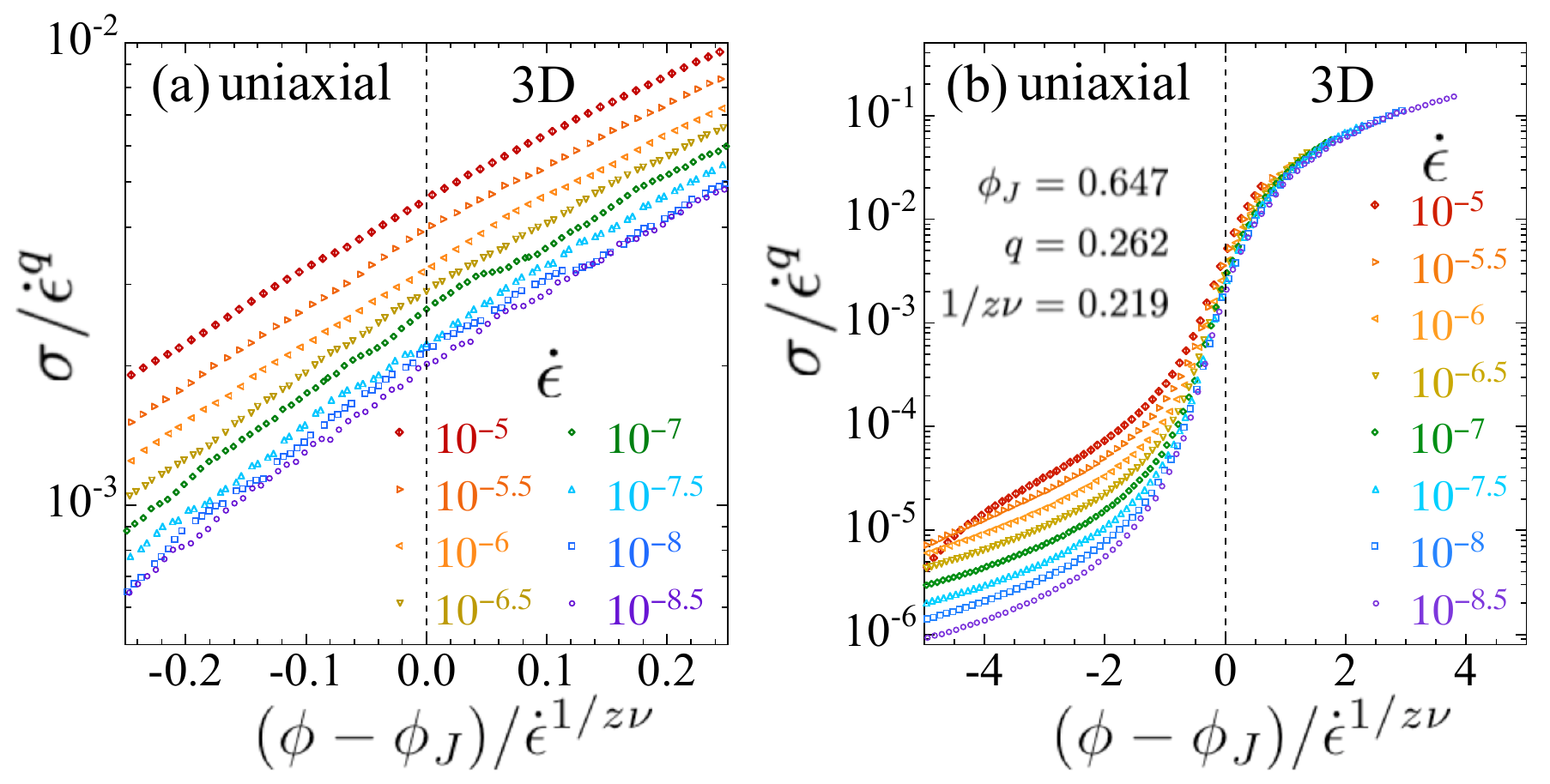}
\caption{{\color{black}Attempted} scaling collapse for the shear stress data shown in Fig.~\ref{f4}, plotting  $\sigma/\dot\epsilon^q$ vs $x=(\phi-\phi_J)/\dot\epsilon^{1/z\nu}$ for uniaxial compression.  The values of $q$, $1/z\nu$ and $\phi_J$ are the same as was obtained from the scaling analysis of pressure $p$, as shown in Fig.~\ref{f3}.  {\color{black}(a) shows a narrow range of $x$, corresponding to the range used in obtaining the fit to pressure $p$ in Fig.~\ref{f3}.  (b) shows a much wider range of $x$ that includes much more of our data.}  We see  {\color{black} that our data appear to collapse} to a common curve as the strain rate $\dot\epsilon$ decreases. However, as compared to what was found for pressure in Fig.~\ref{f3}, here we observe a greater spread {\color{black} of the data at larger $\dot\epsilon$ and } below $\phi_J$ ($x<0$).  This is an indication that corrections to scaling are more significant for $\sigma$ as compared to $p$.
}
\label{f5}
\end{figure}

Since $p$ and $\sigma$ are both parts of the same stress tensor, we expect that they will scale with the same critical parameters.  In Fig.~\ref{f5} we test this assumption by plotting $\sigma/\dot\epsilon^q$ vs $(\phi-\phi_J)/\dot\epsilon^{1/z\nu}$, using the same values for $q$, $1/z\nu$, and $\phi_J$ as were found in Eqs.~(\ref{eexps}) and (\ref{ephiJ}) for the pressure $p$.  We see that this collapse is not {\color{black}nearly} as good as we found for $p$.  A common scaling curve does seem to be emerging as $\dot\epsilon$ decreases, but, {\color{black}compared to what is seen in Fig.~\ref{f3} for the pressure,
here the deviations  are much larger at the larger $\dot\epsilon$, and for $\phi<\phi_J$ below jamming}. We believe that this is due to ``corrections-to-scaling," which come into play whenever one's data is insufficiently close to the critical point, in this case $(\phi_J,\dot\epsilon\to 0)$.  It has previously been found for simple shear-driven jamming that such corrections-to-scaling affect the shear stress
$\sigma$ much more strongly than they do the pressure $p$ \cite{OT2,VagbergOlssonTeitel,Berthier,Rahbari}.

\begin{figure}
\centering
\includegraphics[width=3.4in]{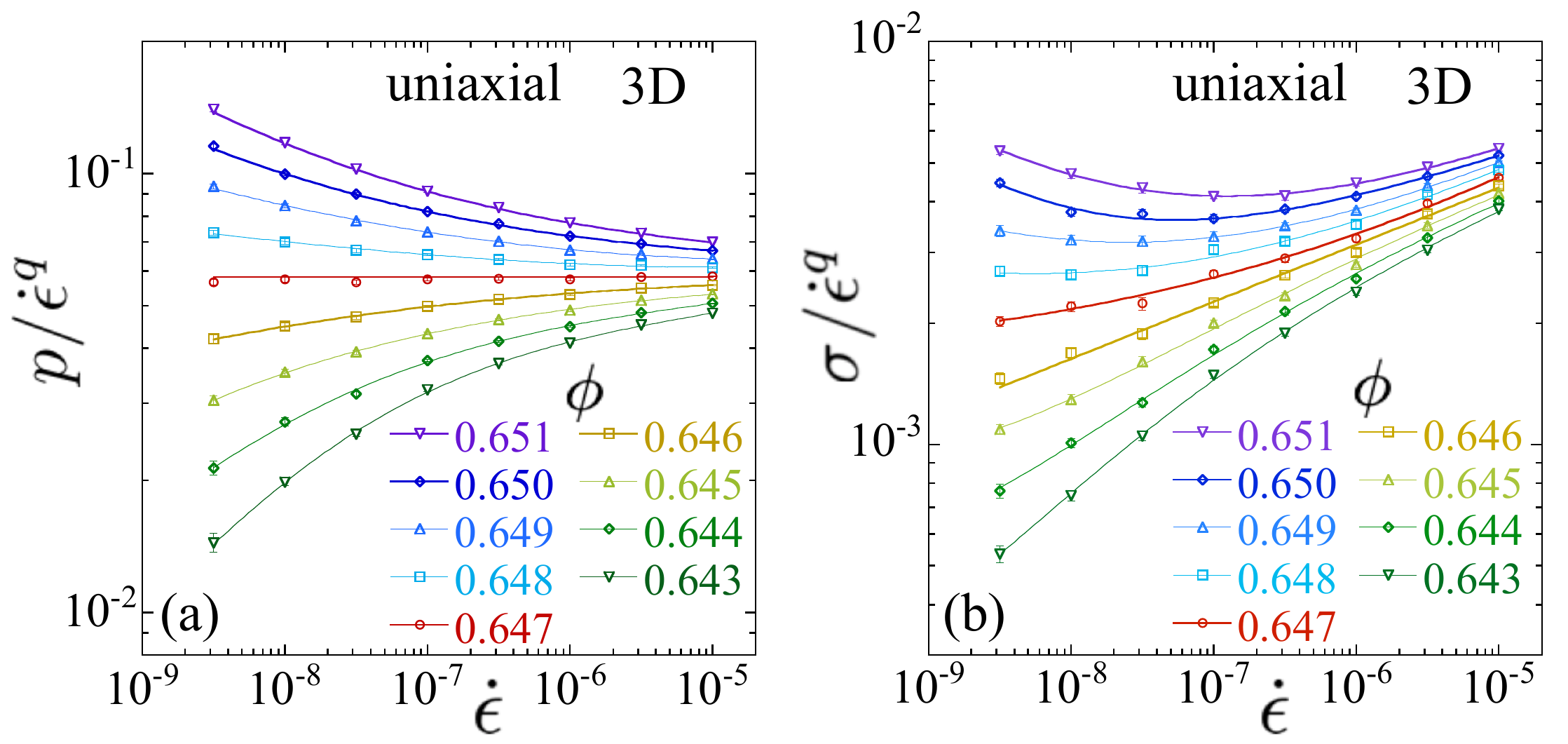}
\caption{Data of Figs.~\ref{f1}(a) and \ref{f4}(a), replotted as (a) scaled pressure $p/\dot\epsilon^q$, and (b) scaled shear stress $\sigma/\dot\epsilon^q$, vs the strain rate $\dot\epsilon$, at different packing fractions $\phi$ about the jamming $\phi_J=0.647$, for uniaxial compression.  At $\phi_J$ (the red data points), the data for $p/\dot\epsilon^q$ is independent of $\dot\epsilon$, consistent with the scaling Eq.~(\ref{escale}).  However, at $\phi_J$, the data for $\sigma/\dot\epsilon^q$ curves upwards, indicating that $\sigma$ is significantly effected by corrections to scaling.  At $\phi_J=0.647$, the solid line in (a) is a fit to a constant, while in (b) it is a fit to the form $c_0 + c_1\dot\epsilon^{\omega/\nu}$.  In (b) the solid line at $\phi=0.646$ is a fit to $c\dot\epsilon^w$.  Solid lines at other values of $\phi$ are simply guides to the eye.
}
\label{f6}
\end{figure}

As another way to see the effect of corrections-to-scaling, in Fig.~\ref{f6}(a)  we plot $p/\dot\epsilon^q$ vs $\dot\epsilon$ for different fixed $\phi$ near $\phi_J$, and in \ref{f6}(b) we similarly plot $\sigma/\dot\epsilon^q$.  From the scaling Eq.~(\ref{escale}) we expect that exactly at $\phi=\phi_J$, $p/\dot\epsilon^q$ will be the constant $f(0)$, independent of $\dot\epsilon$.   In Fig.~\ref{f6}(a) we see just such behavior.  For $\sigma/\dot\epsilon^q$ in Fig.~\ref{f6}(b) however, the data at $\phi=\phi_J=0.647$ are not similarly a constant, but rather increase with increasing $\dot\epsilon$.  

When corrections-to-scaling are important, the scaling equation (\ref{escale}) must be modified to \cite{OT2,VagbergOlssonTeitel},
\begin{equation}
\sigma = \dot\epsilon^q\left[f_1\left(\dfrac{\phi-\phi_J}{\dot\epsilon^{1/z\nu}}\right) +\dot\epsilon^{\omega/z}f_2\left(\dfrac{\phi-\phi_J}{\dot\epsilon^{1/z\nu}}\right) \right],
\label{ects}
\end{equation}
where $\omega>0$ is the correction-to-scaling exponent, coming from the leading irrelevant scaling variable.  
Sufficiently close to the jamming critical point, where $\dot\epsilon\to 0$, the correction term proportional to $f_2$  becomes negligible compared to the leading term $f_1$, and one recovers Eq.~(\ref{escale}).  However when $\dot\epsilon$ is too big, the correction term must be included to characterize the data.
See Ref.~\cite{VagbergOlssonTeitel} for further discussion of corrections-to-scaling in the context of the jamming transition.

We have  tried to fit our data for $\sigma$ to the form of Eq.~(\ref{ects}), approximating both the unknown scaling functions $f_1$ and $f_2$ as in Eq.~(\ref{escalingf}).  Such an approach has been done previously for shear-driven jamming in both two and three dimensions \cite{OT2,Olsson3D,VagbergOlssonTeitel}.  However, in the present case, we find that the degrees of freedom associated with having two unknown scaling functions are too many for us to get reliable results;  our data is not sufficiently accurate to yield stable fits with this method.  We therefore proceed with a simpler approach.  Exactly at $\phi=\phi_J$, Eq.~(\ref{ects}) reduces to,
\begin{equation}
\sigma / \dot\epsilon^q = f_1(0)+\dot\epsilon^{\omega/z} f_2(0),\quad\text{at } \phi=\phi_J.
\label{ects0}
\end{equation}
If we assume the same $\phi_J$ and $q$ as found from our analysis of the pressure $p$, we can then fit the data for $\sigma/\dot\epsilon^q$ at $\phi_J$ to the above form, and from that determine an estimate for $\omega/z$.  Such a fit is shown as the solid line at $\phi_J=0.647$ in Fig.~\ref{f6}(b). The fit has a $\chi^2/n_f=1.55$ and determines the estimate $\omega/z = 0.24\pm 0.02$.

\begin{figure}
\centering
\includegraphics[width=3.4in]{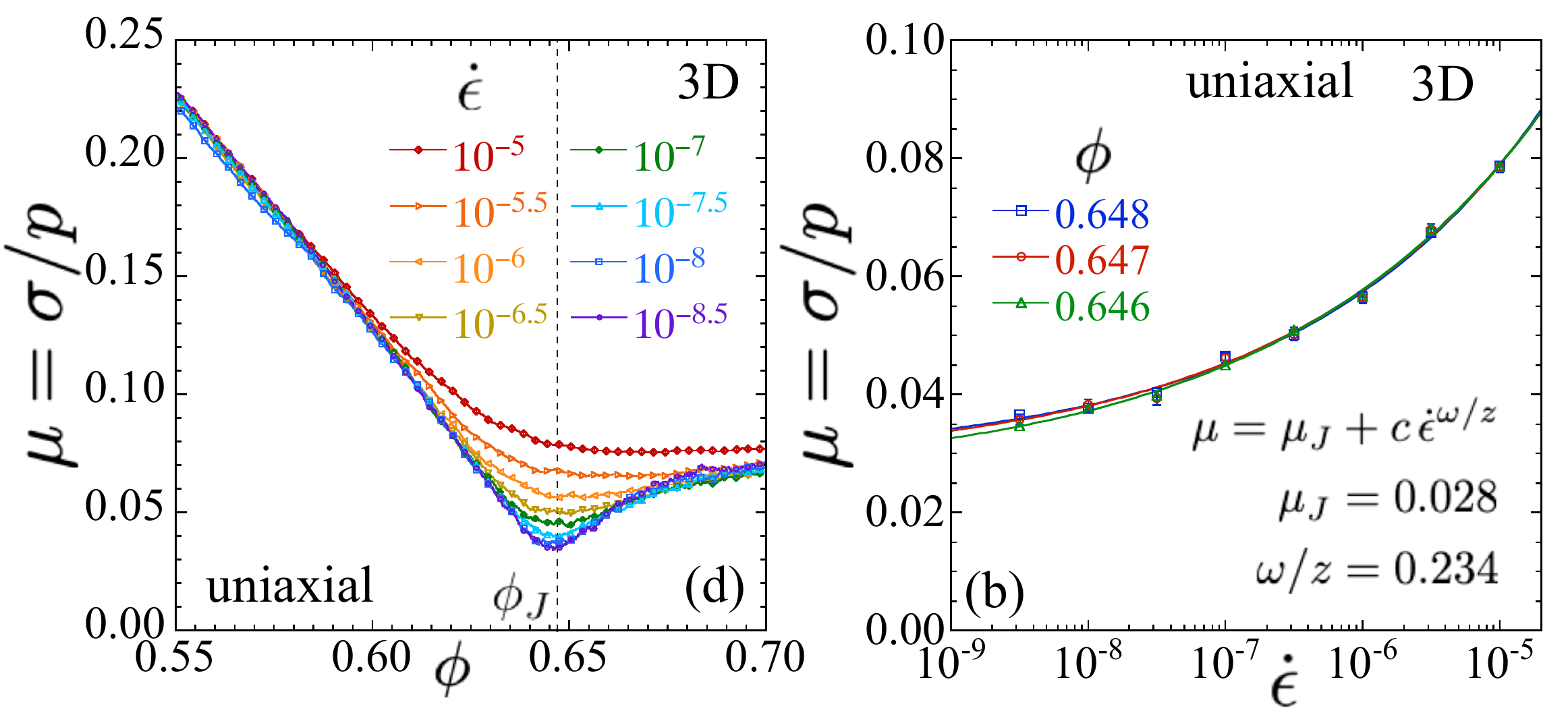}
\caption{(a) The macroscopic friction, $\mu=\sigma/p$, vs packing $\phi$ at different strain rates $\dot\epsilon$ for uniaxial compression.  The vertical dashed line indicates the jamming $\phi_J=0.647$, where $\mu$ has a sharp minimum as $\dot\epsilon\to 0$.  (b) Friction $\mu$ vs $\dot\epsilon$ at $\phi=0.646, 0.647, 0.648$, near the presumed $\phi_J=0.647$.  The solid lines are  fits to $\mu=\mu_J+c\,\dot\epsilon^{\omega/z}$ and at $\phi=0.647$ gives $\mu_J = 0.028\pm 0.002$ and $\omega/z=0.234\pm0.022$.
}
\label{f7}
\end{figure}

Looking solely at Fig.~\ref{f6}(b), one sees that the data at $\phi=0.646$ appears to lie close to a straight line with a finite slope.  The solid line through this data  in \ref{f6}(b) is a fit to a simple power-law, $c\dot\epsilon^w$; the fit has a $\chi^2/n_f=1.80$ and yields $w=0.142\pm 0.003$.  This larger $\chi^2/n_f$,
compared to that obtained when using Eq.~(\ref{ects0}) to describe the data at  $\phi=0.647$, indicates a poorer fit; moreover the data points for the two smallest $\dot\epsilon$ lie above the fitted line.  Yet one might be tempted to think that this behavior indicates that the jamming point for $\sigma$ is $\phi_J^\sigma=0.646$, slightly smaller than $\phi_J^p=0.647$, and that the power-law $q_\sigma=q_p+w =0.404$ is larger than the power-law $q_p=0.262$.

To demonstrate that  this is not the case, and that both $p$ and $\sigma$ do indeed scale with the same exponent $q$, and are characterized by the same jamming $\phi_J$, we consider the macroscopic friction $\mu=\sigma/p$.  If, for example, one had $\phi_J^\sigma<\phi_J^p$, then $\mu=\sigma/p = \eta/\zeta$ would diverge at $\phi_J^\sigma$ as $\dot\epsilon\to 0$.  If one  had $q_\sigma > q_p$, then at $\phi_J$ one would have $\mu\sim \dot\epsilon^{q_\sigma-q_p}$ and $\mu$ would vanish as $\dot\epsilon\to 0$.

In Fig.~\ref{f7}(a) we plot $\mu$ vs $\phi$ for different compression rates $\dot\epsilon$.  We see that, as $\dot\epsilon$ decreases, $\mu$ approaches a finite value at all $\phi$.  Thus we conclude that $\sigma$ and $p$ are both characterized by the same $\phi_J$ and $q$.
In this case, if both $\sigma$ and $p$ obey scaling equations of the form of Eq.~(\ref{ects}), then $\mu$ will have the form,
\begin{equation}
\mu=g_1\left(\dfrac{\phi-\phi_J}{\dot\epsilon^{1/z\nu}}\right) + \dot\epsilon^{\omega/z}g_2\left(\dfrac{\phi-\phi_J}{\dot\epsilon^{1/z\nu}}\right).
\label{escalemu}
\end{equation}
In particular, exactly at $\phi=\phi_J$, the above gives,
\begin{equation}
\mu = \mu_J + c\,\dot\epsilon^{\omega/z},\quad\text{at }\phi=\phi_J,
\end{equation}
where $\mu_J=g_1(0)$ is the quasistatic $\dot\epsilon\to 0$ value of $\mu$ exactly at jamming.  In Fig.~\ref{f7}(b) we plot $\mu$ vs $\dot\epsilon$ for $\phi=0.646$, $0.647$,  $0.648$, near the assumed $\phi_J$, and fit to the above form.  For all three $\phi$ we get an excellent fit with a finite value of $\mu_J$ ranging from 0.026 to 0.029 and an exponent $\omega/z$ ranging from 0.22 to 0.24.  Taking $\phi_J=0.647$ we have  $\mu_J=0.028\pm 0.002$ and $\omega/z = 0.234\pm 0.022$.  This value of $\omega/z$ agrees, within the estimated error, with the value obtained directly from $\sigma$ via Eq.~(\ref{ects0}).  Combined with our earlier result of $1/z\nu = 0.219\pm 0.03$ we then have $\omega\nu=1.07\pm 0.18$, which agrees with the value found previously in two dimensions for simple shearing \cite{OT2}.

Note, as $\dot\epsilon\to 0$,  the shape of $\mu(\phi)$ shown  in Fig.~\ref{f7}(a) is dramatically different from that seen in simple shearing.  In simple shearing, $\mu(\phi)$ is a monotonically decreasing function of $\phi$, with $\mu_J\approx 0.1$ \cite{Vagberg.PRL.2014}.  Here, for uniaxial compression, we see that $\mu(\phi)$ is non-monotonic with a sharp minimum at $\phi_J$, and $\mu_J=0.028$ is roughly a factor 3.5 times smaller than for simple shearing.  However the important point is that, for uniaxial compression, $\mu_J$ remains finite.  For uniaxial compression, configurations at the jamming transition are stress-anisotropic, just as they are for simple shearing, and unlike the stress-isotropic configurations for isotropic compression.

%

\subsection{Scaling of Contact Number}

For isotropically jammed configurations of soft-core, frictionless spheres, such as obtained by quenching random initial configurations, or from isotropic quasistatic compression or decompression, above jamming  the average number of inter-particle contacts per particle $Z$ is found  to obey the scaling law \cite{OHern,Wyart},
\begin{equation}
Z-Z_\mathrm{iso}\sim (\phi-\phi_J)^{1/2},\quad\text{for }\phi>\phi_J,
\label{escaleZ}
\end{equation}
provided rattler particles are removed from the calculation of $Z$.  Exactly at jamming, $Z=Z_\mathrm{iso}=2d$ is the isostatic value, where $d$ is the spatial dimension of the system (so $Z_\mathrm{iso}=6$ in 3D).  A rattler is any particle, in a mechanically stable configuration, which retains at least one unconstrained translational degree of  freedom; rattlers are usually  particles that are trapped within a cage formed by other particles that participate in the system spanning force chain network that characterizes a jammed configuration.
When quenching from random initial configurations, or when compressing or decompressing with energy relaxation between compression steps, one finds $Z=0$ below jamming; particles can avoid all contacts.  Thus, in such cases, $Z$ is said to take a discontinuous jump from zero to $Z_\mathrm{iso}$ at the jamming packing $\phi_J$. 

Recently the same scaling for $Z$ was found for anisotropically jammed frictionless spheres, obtained by the quasistatic shear-jamming of initially unjammed isotropic configurations \cite{Baity,Jin2}.  This indicates that the exponent $1/2$ in Eq.~(\ref{escaleZ}) is universal for both stress-isotropic and stress-anisotropic jamming.  Here we extend the scaling of $Z$ to compression-driven jamming at a finite strain rate $\dot\epsilon$, and show that the scaling equation (\ref{escaleZ}) is recovered in the  $\dot\epsilon\to 0$ limit for both isotropic and uniaxial compression.  
In the following we will denote the average contact number, as computed without rattlers, by $Z$; we will denote the average contact number of all particles, including rattlers, by $Z_\mathrm{all}$.

\begin{figure}
\centering
\includegraphics[width=2.5in]{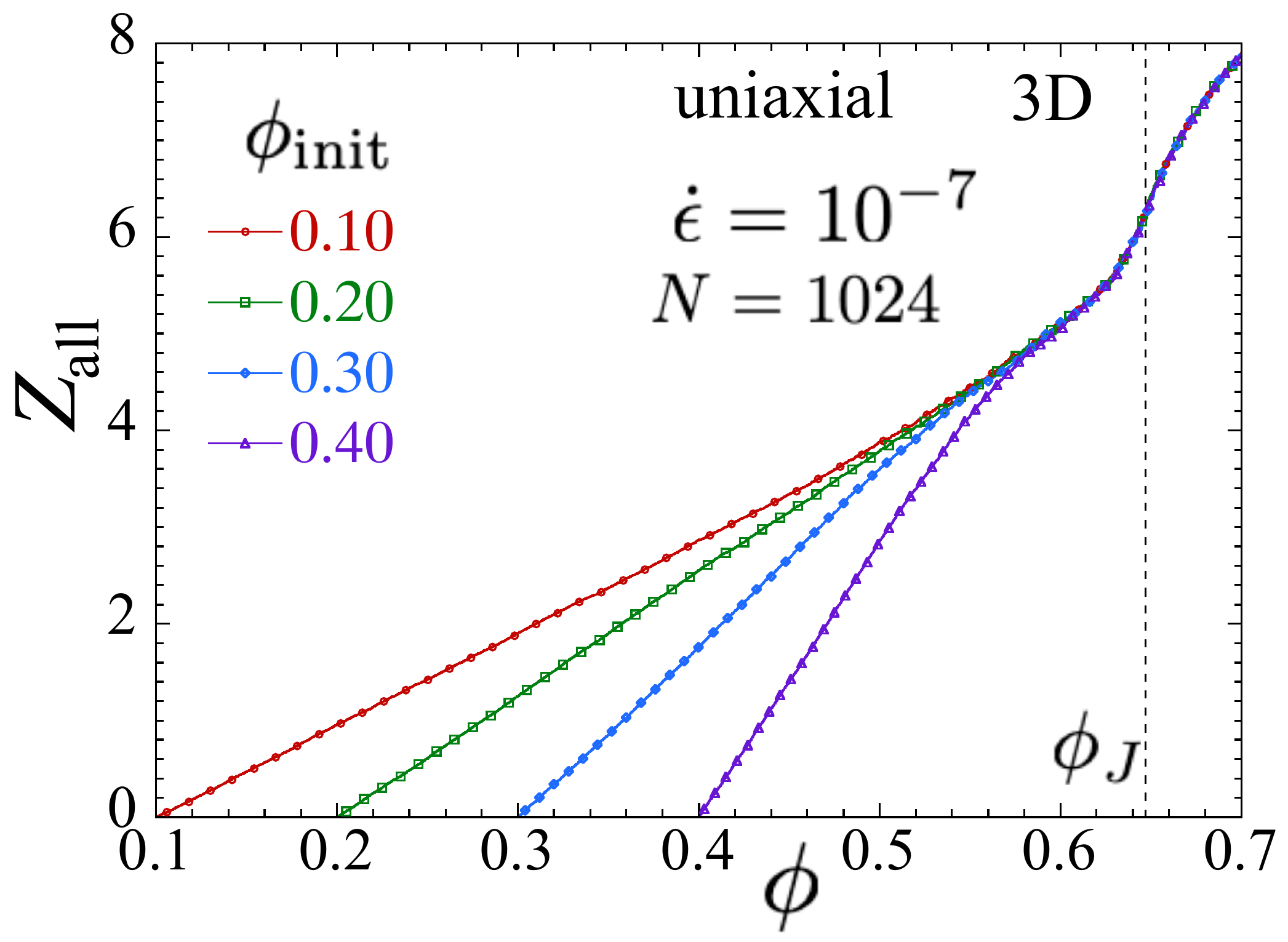}
\caption{Average number of contacts per particle $Z_\mathrm{all}$ vs packing $\phi$ for  uniaxial compression starting from non-overlapping configurations at different $\phi_\mathrm{init}$ at compression strain rate $\dot\epsilon=10^{-7}$.  The system has $N=1024$ particles and results are averaged over 10 independent samples.  Error bars are roughly equal to or smaller than the size of the data symbols.  The vertical dashed line indicates the jamming $\phi_J=0.647$.  Here, rattlers are included in the calculation of $Z_\mathrm{all}$.
}
\label{f8}
\end{figure}

In Fig.~\ref{f8} we plot $Z_\mathrm{all}$ vs $\phi$ for uniaxial compression at the fixed rate $\dot\epsilon=10^{-7}$.  We show results for $N=1024$ particles starting from configurations of non-overlapping but otherwise randomly positioned particles at several different initial packing fractions $\phi_\mathrm{init}$.  Our results at each $\phi_\mathrm{init}$ are averaged over 10 independent compression runs.   
In this figure we show $Z_\mathrm{all}$, which includes rattlers, rather than $Z$, since the notion of a rattler becomes ambiguous at low packings where there are no extended force chains.  In contrast, $Z_\mathrm{all}$ remains  well defined down to $\phi_\mathrm{init}$.

We see that, unlike the methods that involve energy relaxation and give $Z_\mathrm{all}=0$ for $\phi<\phi_J$, here we find a finite $Z_\mathrm{all}$ for all $\phi>\phi_\mathrm{init}$.  We have $Z_\mathrm{all}=0$ at $\phi_\mathrm{init}$ by the definition of how we construct our initial configuration.  The finite $Z_\mathrm{all}$ above $\phi_\mathrm{init}$ is due to our dynamic process of compression, in which  particles push into each other as the system box contracts.  A similar effect of $Z_\mathrm{all}>0$ below $\phi_J$ was seen in simple shearing simulations \cite{Heussinger1}, however there is one important difference between shearing and the present case of compression.  When simple shearing at a fixed rate, the system samples an ensemble of states that becomes independent of the initial configuration, if one shears long enough \cite{Vagberg.PRE.2011}.  When we compress, the effect of the initial configuration, and details of the compression protocol, can strongly effect behavior at low packings.  

Thus, in Fig.~\ref{f8} we see that $Z_\mathrm{all}(\phi)$ at low $\phi$ is clearly different depending upon the particular value of $\phi_\mathrm{init}$ from which we start our compressions. We see that $Z_\mathrm{all}$ rises linearly from zero as $\phi$ increases above $\phi_\mathrm{init}$.   However, we see that $Z_\mathrm{all}(\phi)$ becomes independent of $\phi_\mathrm{init}$ once $\phi\gtrsim 0.58$.  
Thus near jamming, our calculation of the average contact number becomes independent of the particular $\phi_\mathrm{init}$ from which we begin our compression.
A similar independence of the pressure $p$ on $\phi_\mathrm{init}$, at large $\phi$ near and above jamming, was found in \cite{supp}.

\begin{figure}
\centering
\includegraphics[width=3.4in]{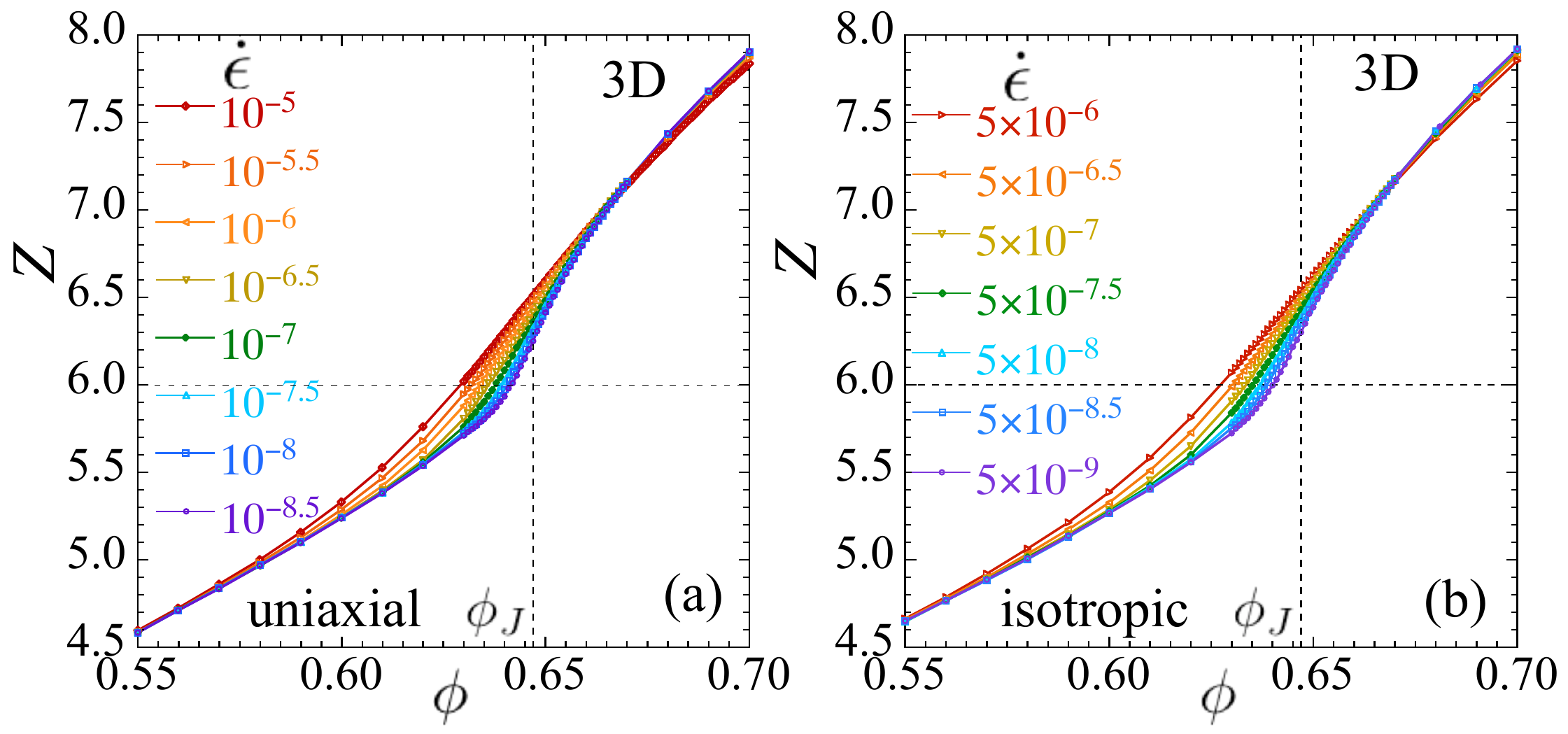}
\caption{Average number of contacts per particle $Z$ vs packing $\phi$ at different strain rates $\dot\epsilon$ for (a) uniaxial compression and (b) isotropic compression.  Rattlers have been removed from the system before computing $Z$.  The vertical dashed lines indicate the jamming $\phi_J=0.647$.  The horizontal dashed lines indicate the isostatic value $Z_\mathrm{iso}=6$.  The system has $N= 32768$ particles and starts from $\phi_\mathrm{init}=0.2$.
Error bars are roughly equal to or smaller than the size of the data symbols.
}
\label{f9}
\end{figure}

We now focus on the contact number $Z$, obtained after  first removing all rattlers from the system, at larger packings near jamming.
We define a rattler as any particle which has fewer than four contacts.  We recursively loop through the system, removing rattlers until no further rattlers are found.
In Fig.~\ref{f9} we plot $Z$ vs $\phi$ for different strain rates $\dot\epsilon$, for a system of $N=32768$ particles.  
Fig.~\ref{f9}(a) shows our results for uniaxial compression while \ref{f9}(b) is for isotropic compression.  In both cases the compression starts from $\phi_\mathrm{init}=0.2$.  We see that $Z$ approaches a limiting curve as $\dot\epsilon$ decreases, and that the dependence of $Z$ on $\dot\epsilon$  is only readily apparent in the vicinity of $\phi_J$.

We wish to generalize the scaling equation (\ref{escaleZ}) to include the strain rate $\dot\epsilon$, in a similar manner to Eq.~(\ref{escale}).   The difficulty is that, unlike $p$ or $\sigma$ which vanish below $\phi_J$, the contact number $Z$ is finite below $\phi_J$.  We take a phenomenological approach and assume that the contacts below $\phi_J$, that arise from our protocol of compressing, constitute a smooth non-singular contribution to $Z$ that must be subtracted to get the critical part that scales.  Generalizing to finite strain rates $\dot\epsilon$, we therefore posit the scaling form,
\begin{equation}
Z-Z_\mathrm{ns}(\phi)=\dot\epsilon^\kappa h\left(\dfrac{\phi-\phi_J}{\dot\epsilon^{1/z\nu}}\right),
\label{escaledZ}
\end{equation}
where $Z_\mathrm{ns}(\phi)$ is the non-singular part, 
and $Z_\mathrm{ns}(\phi_J)=Z_\mathrm{iso}$.  
For $\phi<\phi_J$, we expect $Z-Z_\mathrm{ns}\to 0$ as $\dot\epsilon\to 0$; hence, below $\phi_J$,  $Z_\mathrm{ns}$ is just the $\lim_{\dot\epsilon\to 0}Z$.  For $\phi>\phi_J$, we expect that $Z-Z_\mathrm{ns}$ approaches a finite constant as $\dot\epsilon\to 0$.  We therefore expect $h(x)\sim x^{\kappa z\nu}$ as $x\to +\infty$, and so as $\dot\epsilon\to 0$,  $Z-Z_\mathrm{ns}\sim(\phi-\phi_J)^{\kappa z\nu}$.  For this to agree with Eq.~(\ref{escaleZ}), we must then have $\kappa z\nu=1/2$, or $\kappa=1/2z\nu$.

\begin{figure}
\centering
\includegraphics[width=3.4in]{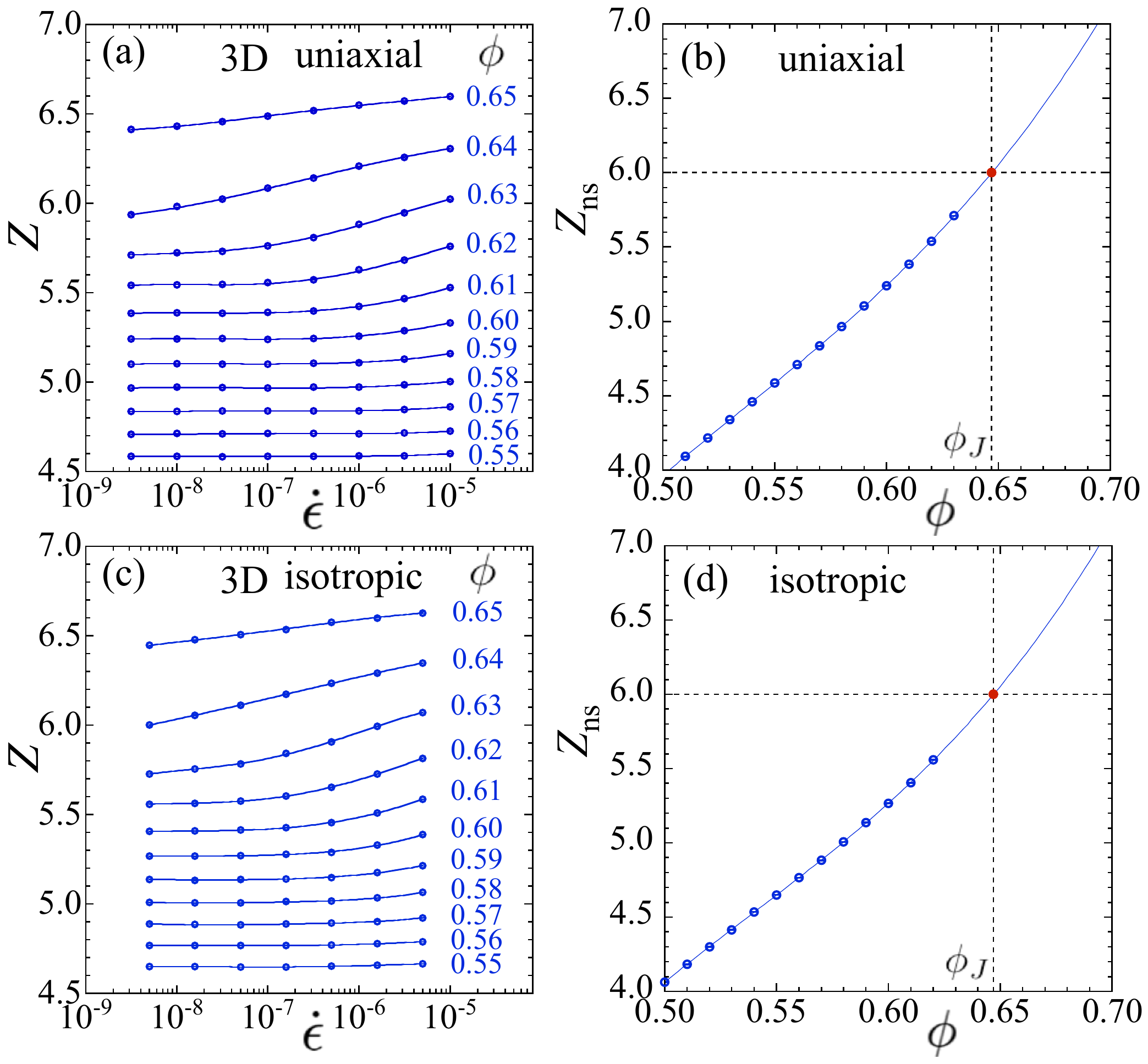}
\caption{Average contact number $Z$ vs strain rate $\dot\epsilon$, for different packings $\phi$ below and just above the jamming $\phi_J=0.647$, for (a) uniaxial  and (c) isotropic compression.  Limiting value $\lim_{\dot\epsilon\to 0} Z$ defines $Z_\mathrm{ns}(\phi)$, the non-singular contribution to $Z$ for $\phi$ below $\phi_J$ for (b) uniaxial  and (d) isotropic compression.  In (b) and (d) 
the solid line is a fit to a cubic polynomial of the form given in Eq.~(\ref{pfit}).  The vertical dashed lines in (b) and (d) locate the jamming $\phi_J$, while the horizontal dashed lines locate the isostatic value $Z_\mathrm{iso}=6$.  The red dots in (b) and (d) indicate the isostatic point $(\phi_J,Z_\mathrm{iso})$.
}
\label{f10}
\end{figure}

To determine $Z_\mathrm{ns}(\phi)$ for uniaxial compression, we plot $Z$ vs $\dot\epsilon$ at various different packings $\phi$   in Fig.~\ref{f10}(a).  When the values of $Z$ at the two lowest rates $\dot\epsilon$ are equal, within the estimated errors, we take that value as the $\dot\epsilon\to 0$ limit $Z_\mathrm{ns}(\phi)$.  With this approach we can  obtain $Z_\mathrm{ns}(\phi)$ for packings $\phi$ up to  $\phi=0.63$. To analytically continue $Z_\mathrm{ns}$ up to and above $\phi_J$, we then fit the data points so obtained to an $n$th order polynomial, 
\begin{equation}
Z_\mathrm{ns}(\phi)=6-\sum_{m=1}^n(\phi_J-\phi)^m.
\label{pfit}
\end{equation}
The form of the polynomial above guarantees that $Z_\mathrm{ns}$ passes through the isostatic point, 
$Z_\mathrm{ns}(\phi_J)=Z_\mathrm{iso}=6$.
Using our previously determined $\phi_J=0.647$, we find good results using a cubic polynomial with $n=3$.
In Fig.~\ref{f10}(b) we plot the data points for $Z_\mathrm{ns}$ vs $\phi$ and show results for this cubit fit.
In Figs.~\ref{f10}(c) and \ref{f10}(d) we show the corresponding plots for isotropic compression.  For this case we are able to obtain $Z_\mathrm{ns}(\phi)$ only up to $\phi=0.62$; a cubic polynomial is again used to extrapolate $Z_\mathrm{ns}$ to larger $\phi$.


\begin{figure}
\centering
\includegraphics[width=3.4in]{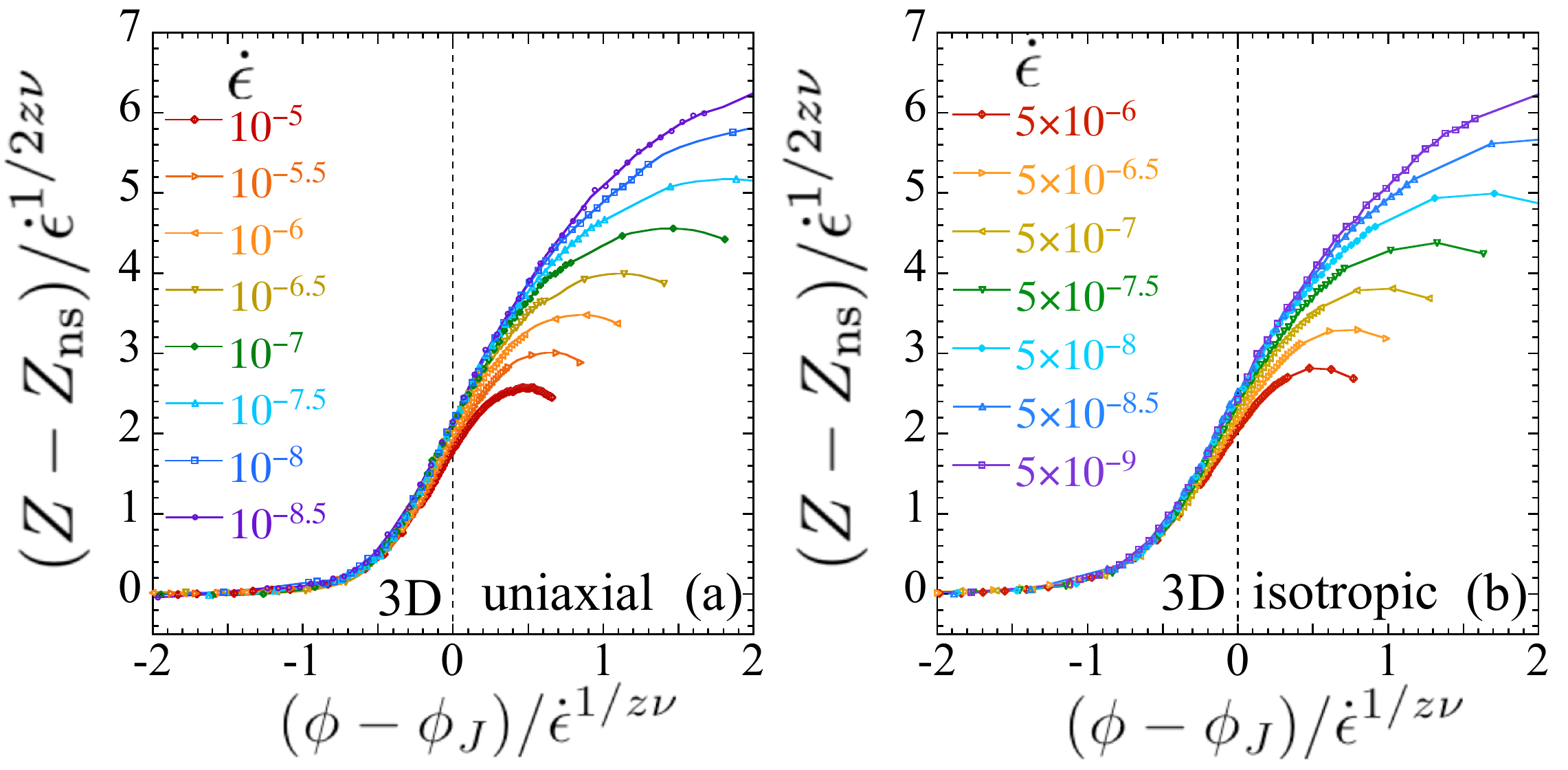}
\caption{Scaling collapse for the singular part of the average contact number $Z-Z_\mathrm{ns}$, plotting $(Z-Z_\mathrm{ns})/\dot\epsilon^{1/2z\nu}$ vs $x=(\phi-\phi_J)/\dot\epsilon^{1/z\nu}$ at different strain rates $\dot\epsilon$ for (a) uniaxial and (b) isotropic compression. 
}
\label{f11}
\end{figure}

Using the above determined $Z_\mathrm{ns}$,  setting $\kappa=1/2z\nu$, and using the same values of $\phi_J$ and $1/z\nu$ given in Eq.~(\ref{eexps}) that were found from our fits to the pressure $p$ ,  in Fig.~\ref{f11} we plot $(Z-Z_\mathrm{ns})/\dot\epsilon^{1/2z\nu}$ vs $x=(\phi-\phi_J)/\dot\epsilon^{1/z\nu}$, so as to test the scaling prediction of Eq.~(\ref{escaledZ}).  Fig.~\ref{f11}(a) shows the scaling collapse for uniaxial compression, while \ref{f11}(b) is for isotropic compression.  We see an excellent data collapse for $x<0$, below jamming.  The collapse remains good for $x>0$, above jamming, with the curves of different $\dot\epsilon$ peeling off from the limiting $\dot\epsilon\to 0$ curve at successively smaller values of $x$ as $\dot\epsilon$ increases.  The departure from a common scaling curve above $\phi_J$ as $\dot\epsilon$ increases might be due to corrections-to-scaling.   We note that the largest $x$ data point on each $\dot\epsilon$ curve corresponds to the packing fraction $\phi=0.70$, which is sufficiently above $\phi_J=0.647$  that we would not expect it to be described by just the leading scaling term.  However, we believe it is more likely that the departure from a common scaling curve as $x$ increases above zero is due primarily to the failure of our predicted $Z_\mathrm{ns}(\phi)$, determined solely from data below $\phi_J$, to remain accurate as we go much above $\phi_J$.

The good collapses we see in Fig.~\ref{f11} thus show that the effect of a finite compression rate on the contact number $Z$ is governed by the same scaling  
variable $(\phi-\phi_J)/\dot\epsilon^{1/z\nu}$ that was found for the pressure $p$.  Moreover, the quasistatic limit is correctly described by the exponent $1/2$, as in Eq.~(\ref{escaleZ}).   As with the scaling of pressure $p$, we see that the critical exponents characterizing the contact number $Z$ are universal, being the same for stress-anisotropic jamming (uniaxial compression) as for stress-isotropic jamming (isotropic compression).

\section{Discussion}
\label{Discuss}

Our results strongly argue that, in three dimensions, stress-isotropic and stress-anisotropic  jamming are in the same critical universality class, not only for static structural properties, but also for dynamic properties governed by a diverging critical time scale.  For both isotropic compression, where  configurations have an isotropic stress tensor with $\mu=\sigma/p=0$, and for uniaxial compression, where   configurations have an anisotropic stress tensor with a finite $\mu=\sigma/p>0$, we find that the bulk viscosity diverges as $\zeta=p/\dot\epsilon\sim (\phi_J-\phi)^{-\beta}$, with a common $\beta=3.36\pm 0.09$.  Our discussion of the scaling of the shear stress $\sigma$, and the macroscopic friction $\mu$, suggests that the shear viscosity in uniaxial compression, $\eta=\sigma/\dot\epsilon$, diverges similarly, though with strong corrections-to-scaling characterized by the correction exponent $\omega/z=0.24\pm0.02$.

We now compare our results with other simulations in the literature.  We consider only works on three dimensional systems, with the same particle size-dispersity we use here, and with a similar viscously overdamped dynamics.
To compare  results, two key quantities are the number of particles  $N$ in the system, and the range of packing fractions $\phi_\mathrm{min}\le\phi\le\phi_\mathrm{max}$ that are used in fitting to the numerical data.  We therefore define $\delta\phi_\mathrm{max}/\phi_J = (\phi_J-\phi_\mathrm{min})/\phi_J$, which measures the relative distance to jamming of the data that is farthest from $\phi_J$.  Recall, critical scaling holds only asymptotically close to the critical point, so the smaller is $\delta\phi_\mathrm{max}/\phi_J$, the more likely one is to be in this asymptotic critical region.  For the fits that determined the values of our critical exponents given above, we used $N=32768$ particles and  a data window of $\delta\phi_\mathrm{max}/\phi_J\approx 0.008$.

Recent 3D simulations by Ikeda and Hukushima \cite{HIkeda2S} have computed a quantity analogous to the bulk viscosity by considering particle displacements under quasistatic isotropic compression. Using a finite-size scaling analysis for bidisperse systems with $N\le 4096$ and a relatively large data window of $\delta\phi_\mathrm{max}/\phi_J=0.15$, they claimed $\beta= 2.7$.  These are the only other simulations we are aware of that  address the divergence of the bulk viscosity under compression.

We can also compare our results to 
those in the literature for stress-anisotropic simple shear-driven jamming.  For a simple-shear strain rate $\dot\gamma$, we  define the pressure analog of shear viscosity as $\eta_p=p/\dot\gamma\sim(\phi_J-\phi)^{-\beta}$.  
Lerner et al. \cite{Lerner}, simulated 1000  hard-core spheres and obtained $\beta=2.63$ by fitting over a data window with $\delta\phi_\mathrm{max}/\phi_J\approx 0.062$.
DeGiuli et al. \cite{DeGiuli}, also using 1000  hard-core spheres, found $\beta=2.78$ by fitting over a data window with $\delta\phi_\mathrm{max}/\phi_J\approx 0.043$.
Kawasaki et al. \cite{Berthier} simulated up to 10000  soft-core spheres and found $\beta=2.56$ by making an extrapolation to the $N\to\infty$ and $\dot\gamma\to 0$ hard-core limit; they used a data window with $\delta\phi_\mathrm{max}/\phi_J\approx 0.116$.  Thus we can note that including data that is further away from $\phi_J$ in one's fit, i.e. using a larger $\delta\phi_\mathrm{max}/\phi_J$, seems to result in smaller  values of $\beta$.  We thus believe that the fits in these works include data that is too far from $\phi_J$ to be in the asymptotic critical region, and hence the resulting values of $\beta$ do not reflect the correct asymptotic value.

More recent simulations by Olsson \cite{Olsson3D} with  65536 soft-core spheres, using a scaling analysis similar to that described here with a data window $\delta\phi_\mathrm{max}/\phi_J=0.026$ and $\dot\gamma_\mathrm{max}=10^{-5}$, but explicitly including  corrections-to-scaling in the analysis of both $p$ and $\sigma$, find $\beta=3.8\pm 0.1$, $y=1.16\pm 0.01$, and $\omega/z=0.30\pm 0.02$ (errors cited here are one standard deviation estimated error).  These values are roughly $3-4$ standard deviations away from the values we find in the present work for uniaxial compression.  It could be that the complications associated with including corrections-to-scaling have led to systematic errors resulting in Olsson finding a larger $\beta$ than the correct value; or it could be that the absence of corrections-to-scaling in our analysis of $p$ has led to systematic errors resulting in our finding a smaller $\beta$.  However, we cannot rule out the possibility that simple-shearing and uniaxial compression, though both producing states with anisotropic stress, might be in different universality classes.   Simple-shearing creates  ensembles of configurations that are statistically independent of each other at each value of  $\dot\gamma$ and $\phi$.  Compression, however,  creates ensembles in which configurations for the same $\dot\epsilon$, but different $\phi$, are necessarily correlated.

Finally we can compare our results against theoretical predictions.  DeGiuli et al. \cite{DeGiuli} and D{\"u}ring et al. \cite{During}, treating hard-core spheres below jamming, have considered the relation between $\eta_p{\color{black}=p/\dot\gamma}$ and the deviation of the average contact number from isostaticity, $\eta_p\sim (Z_\mathrm{iso}-Z)^{-\beta^\prime}$.  Using marginal stability arguments, they have proposed that the exponent $\beta^\prime$ can be expressed as $\beta^\prime=(4+2\theta)/(1+\theta)$, where $\theta$ describes the algebraic distribution of the magnitudes of the contact forces $f_{ij}\equiv |\mathbf{f}_{ij}^\mathrm{el}|$ that participate in the extended force network exactly at $\phi_J$, $\mathcal{P}( f_{ij} )\sim f_{ij}^\theta$.  
Recently, Ikeda has presented a  calculation \cite{HIkeda1} of the critical relaxation time $\tau$  from the dynamical matrix of jammed configurations, and  found  $\tau\sim (Z_\mathrm{iso}-Z)^{-\beta^\prime}$ with $\beta^\prime$ related to $\theta$ by the same relationship as above.  
Such a relation between $\beta^\prime$ and $\theta$ would provide a connection between structural and dynamic properties.

An infinite dimensional mean-field theory of the isotropic jamming transition by Charbonneau et al. \cite{Charb2S,Charb1S} has computed the value $\theta=0.423$.  Numerical simulations  \cite{DeGiuli2S,Charb0S} of thermalized and athermal spheres in finite dimensions $d=2,3,4$ have found values of $\theta$ consistent with this prediction.  It has been argued \cite{Wyart,Wyart3S,GoodrichS,CharbonneauS,Goodrich2S}  that the upper critical dimension for jamming may be $d=2$, and if so mean-field critical exponents would apply for all dimensions $d>2$.  Moreover, a common value for $\theta$ was found  \cite{Jin2, Urbani} for thermalized hard-core spheres in both stress-isotropic and stress-anisotropic jammed configurations.
These results thus suggest a common universality for isotropic and anisotropic jamming, and using the above value of $\theta$ one finds $\beta^\prime=3.41$.

Tests of this dependence of $\eta_p$ (or equivalently the relaxation time $\tau$) on $Z_\mathrm{iso}-Z$ have been made in 3D simulations of hard-core spheres, or by taking the $\dot\gamma\to 0$ limit of soft-core spheres.
Measuring $\eta_p$ for sheared  hard-core spheres,  Lerner et al. \cite{Lerner} found the value $\beta^\prime=2.94$ for $N=1000$.  Similar simulations by DeGiuli et al. \cite{DeGiuli} found $\beta^\prime=3.33$ for $N=1000$.  Olsson  \cite{Olsson3D} measured the long time relaxation $\tau$ of $N=65538$  soft-core particles, using initial configurations sampled from steady-state shearing at a finite shear strain rate $\gdot$, and found $\beta^\prime=3.7$.
Ikeda et al. \cite{Ikeda}  similarly measured  the long time relaxation $\tau$ for $N=3000$ particles.  For both initial random isotropic configurations and configurations sampled from shearing  at a finite $\gdot$, they found  all their data to give a common value $\beta^\prime=3.2$.  

To convert this prediction for $\beta^\prime$ into the exponent $\beta$ that we measure in our present work, we need to know how the contact number difference $(Z_\mathrm{iso}- Z)$ scales with the distance in packing below jamming $(\phi_J-\phi)$,
\begin{equation}
Z_\mathrm{iso}-Z\sim (\phi_J-\phi)^u\quad\text{for }\phi<\phi_J.
\end{equation}
Then we will have $\beta=u\beta^\prime$.
Early shearing simulations by Heussinger and Barrat \cite{Heussinger1} claimed $u=1$, and hence $\beta=\beta^\prime$.  However analytic arguments by DeGiuli et al. \cite{DeGiuli} claimed $u=(2+2\theta)/(3+\theta)=0.83$, thus predicting $\beta=2.84$.  As we have discussed above, several numerical works have claimed values of $\beta$ in this neighborhood; however, as we have highlighted, these values seem dependent on the window of data $\delta\phi_\mathrm{max}/\phi_J$ used in the scaling fit, with $\beta$ increasing as $\delta\phi_\mathrm{max}/\phi_J$ decreases.
Indeed, in DeGiuli et al.  \cite{DeGiuli} the fit of their numerical data for $p/\dot\gamma$ vs $\phi_J-\phi$, that is used to determine their value $\beta=2.78$, and the fit of $p/\dot\gamma$ vs $Z_\mathrm{iso}-Z$, that is used to determine their value $\beta^\prime=3.33$, seem to use  almost non-overlaping ranges of $p/\dot\gamma$, with the latter two orders of magnitude closer to the critical point than the former  (see their Fig.~5 and note that their $\mathcal{J}\propto \dot\gamma/p$).

In contrast, Olsson, has done a careful numerical analysis \cite{OlssonRelax,Olsson3D} in both 2D and 3D  that strongly argues $u=1$.  
Our own result in Sec.~\ref{Results}.C, that shows that $Z$ obeys a good scaling collapse when we assume that the background $Z_\mathrm{ns}$ is a non-singular function of the packing as $\phi$ varies through $\phi_J$, is consistent with this conclusion.
If this is the case, then $\beta=\beta^\prime$ and our result $\beta=3.36\pm  0.09$ would be in very good agreement with the marginal stability prediction of $\beta^\prime=3.41$.


\begin{acknowledgments}
We thank P. Olsson for helpful discussions.
This work was supported by National Science Foundation Grant No. DMR-1809318. Computations were carried out at the Center for Integrated Research Computing at the University of Rochester. 
\end{acknowledgments}



\bibliographystyle{apsrev4-1}

\end{document}